\documentclass[a4paper,fleqn]{cas-sc} 

\usepackage[authoryear]{natbib}

\usepackage{graphicx}	
\usepackage{amsmath}	
\usepackage{pdflscape}
\usepackage{subcaption}
\usepackage{gensymb}
\usepackage{array,multirow}
\usepackage{setspace} 
\usepackage{lineno} 

\def\tsc#1{\csdef{#1}{\textsc{\lowercase{#1}}\xspace}}
\tsc{WGM}
\tsc{QE}
\tsc{EP}
\tsc{PMS}
\tsc{BEC}
\tsc{DE}

\begin{document}
\let\WriteBookmarks\relax
\def\floatpagepagefraction{1}
\def\textpagefraction{.001}

\shorttitle{Spectro-photometry of Phobos simulants I.}

\shortauthors{A. Wargnier et~al.}

\title[mode = title]{Spectro-photometry of Phobos simulants}
\title[mode=sub]{I. Detectability of hydrated minerals and organic bands}

\author[1,2]{Antonin Wargnier}[type=editor,
                        auid=000,bioid=1,
                        orcid=0000-0001-7511-2910]

\cormark[1]


\ead{antonin.wargnier@obspm.fr}


\credit{Conceptualization, Data curation, Formal analysis, Investigation, Methodology, Resources, Writing - original draft}

\affiliation[1]{organization={LESIA, Observatoire de Paris, Université PSL, CNRS, Université Paris Cité, Sorbonne Université},
    addressline={5 place Jules Janssen}, 
    city={Meudon},
    postcode={92195}, 
    country={France}}
\affiliation[2]{organization={LATMOS, CNRS, Université Versailles St-Quentin, Université Paris-Saclay, Sorbonne Université},
    addressline={11 Bvd d'Alembert}, 
    city={Guyancourt},
    postcode={F-78280}, 
    country={France}}

\author[2,1]{Thomas Gautier}
\credit{Conceptualization, Funding acquisition, Methodology, Supervision, Validation, Resources, Writing - review \& editing}

\author[1]{Alain Doressoundiram}
\credit{Conceptualization, Project administration, Writing - review \& editing}

\author[1,3]{Giovanni Poggiali}
\credit{Conceptualization, Methodology, Resources, Validation, Writing - review \& editing}
\affiliation[3]{organization={INAF-Astrophysical Observatory of Arcetri},
    addressline={largo E. Fermi n.5}, 
    city={Firenze},
    postcode={I-50125}, 
    country={Italy}}

\author[4]{Pierre Beck}
\credit{Conceptualization, Investigation, Resources, Writing - review \& editing}
\affiliation[4]{organization={Univ. Grenoble Alpes, CNRS, IPAG},
    city={Grenoble},
    postcode={F-38000}, 
    country={France}}

\author[4]{Olivier Poch}
\credit{Investigation, Writing - review \& editing}

\author[4]{Eric Quirico}
\credit{Investigation, Writing - review \& editing}

\author[5]{Tomoki Nakamura}
\credit{Conceptualization, Writing - review \& editing}
\affiliation[5]{organization={Department of Earth Science, Tohoku University},
    state={Sendai},
    country={Japan}}
    
\author[6]{Hideaki Miyamoto}
\credit{Resources, Writing - review \& editing}
\affiliation[6]{organization={Department of Systems Innovation, University of Tokyo},
    addressline={7-3-1 Hongo, Bunkyo-ku}, 
    city={Tokyo},
    country={Japan}}
    
\author[7]{Shingo Kameda}
\credit{Resources, Writing - review \& editing}
\affiliation[7]{organization={Rikkyo University},
    city={Tokyo},
    country={Japan}}
  
\author[8]{Pedro H. Hasselmann}
\credit{Formal analysis, Writing - review \& editing}
\affiliation[8]{organization={INAF - Osservatorio di Roma},
   city={Monte Porzio Catone},
   country={Italy}}
   
\author[9]{Nathalie Ruscassier}
\credit{Investigation, Writing - review \& editing}
\author[9]{Arnaud Buch}
\credit{Resources, Writing - review \& editing}
\affiliation[9]{organization={Laboratoire Génie des Procédés et Matériaux, CentraleSupélec, Université Paris-Saclay},
    city={Gif-sur-Yvette},
    country={France}}

\author[1,10]{Sonia Fornasier}
\credit{Writing - review \& editing}
\affiliation[10]{organization={Institut Universitaire de France (IUF)},
    addressline={1 rue Descartes}, 
    city={Paris Cedex 05},
    postcode={F-75231}, 
    country={France}}
    
\author[1]{Maria Antonietta Barucci}
\credit{Writing - review \& editing}

\cortext[cor1]{Corresponding author}

\begin{abstract}
Previous Mars Reconnaissance Orbiter and Mars Express observations of Phobos and Deimos, the moons of Mars, have improved our understanding of these small bodies. However, their formation and composition remain poorly constrained. Physical and spectral properties suggest that Phobos may be a weakly thermal-altered captured asteroid but the dynamical properties of the martian system suggest a formation by giant collision similar to the Earth moon. In 2027, the JAXA's MMX mission aims to address these outstanding questions. \\
We undertook measurements with a new simulant called OPPS (Observatory of Paris Phobos Simulant) which closely matches Phobos spectra in the visible to the mid-infrared range. The simulant was synthesized using a mixture of olivine, saponite, anthracite, and coal. \\
Since observation geometry is a crucial aspect of planetary surface remote sensing exploration, we evaluated the parameters obtained by modeling the phase curves -- obtained through laboratory measurements -- of two different Phobos simulants (UTPS-TB and OPPS) using Hapke IMSA model. Our results show that the photometric properties of Phobos simulants are not fully consistent with those of Tagish Lake, Allende, or the NWA 4766 shergottite. The single scattering albedo (SSA) of the UTPS-TB is comparable to that of Allende, and the OPPS SSA is similar to that of Tagish Lake. However, other properties, such as opposition effect parameters are more complicated to relate to meteorites.\\
We also investigated the detection of volatiles/organic compounds and hydrated minerals, as the presence of such components is expected on Phobos in the hypothesis of a captured primitive asteroid. To investigate their detectability, we examined the variability of the 3.28 µm and 3.42 µm absorption bands related to aliphatic/aromatic carbon (as a proxy of organic material), as well as the 2.7 µm O-H feature in a Phobos laboratory spectroscopic simulant. The results indicate that a significant amount of organic compounds is required for the detection of C-H bands at 3.4 µm. The bands at 3.28 and 3.42 µm are faint (less than 2\%) when $\sim$3 wt.\% of organic compounds are present in the simulant and are likely undetectable by the MIRS spectrometer onboard the MMX mission. When the concentration of aliphatic and aromatic compounds is increased to 6 wt.\%, a positive detection starts to become more plausible using remote sensing infrared spectroscopy. In contrast, the 2.7 µm absorption band, due to hydrated minerals, is much deeper and easier to detect than C-H organic features at the same concentration levels. The feature is still clearly detectable even when the simulant contains only 3 vol.\% of phyllosilicates, corresponding to 0.7 wt.\% OH groups.\\
Posing limits on detectability of some possible key components of Phobos surface will be pivotal to prepare and interpret future observations of the MIRS spectrometer as well as TENGOO and OROCHI cameras onboard MMX mission.
\end{abstract}

\begin{highlights}
\item Phyllosilicates were found to match well the mid-infrared spectrum of Phobos
\item Organics and hydrated minerals will be detectable by the MIRS instrument onboard MMX
\item Photometric properties of Phobos simulants shows differences compared to carbonaceous chondrites or shergottite 
\end{highlights}

\begin{keywords}
Martian satellites \sep Infrared spectroscopy \sep Spectrophotometry \sep Laboratory astrophysics
\end{keywords}

\maketitle

\protect\doublespacing

\section{Introduction} \label{sect:intro}

Phobos and Deimos, the two martian satellites, are the targets of the JAXA Martian Moon eXploration sample return mission \citep{Kuramoto_2022}. The origin of the two moons is still not well understood and several hypotheses for their formation have been suggested. According with the in-situ scenario, a giant impact between Mars and a protoplanet would have formed Phobos and Deimos by accretion from a disk of debris originated by Mars and the impactor (e.g., \citealt{Craddock_1994, Craddock_2011, Rosenblatt_2011, Hyodo_2017}). In contrast, the asteroid captured hypothesis proposed that the two martian moons were primitive asteroids captured in Mars orbit \citep{Pang_1978, Pajola_2013}. More recently, the hypothesis of a captured comet has also been proposed to explain the origin of Phobos \citep{Fornasier_2023}.\par
The detection, or not, of certain specific absorption features on the surface of Phobos may be of major importance to decipher its origin. In particular, organics and hydrated minerals are expected to be present on Phobos surface in the case of the captured asteroid hypothesis. In this theory, to match the spectroscopic observations, Phobos would be a primitive asteroid in the D-type class according to Bus - DeMeo taxonomy \citep{DeMeo_2009}. Nevertheless, organics and hydrated minerals have never been unambiguously detected on Phobos (and Deimos) surface since all the past visible and near-infrared observations have shown a red spectrum with a gap in the 2.7 µm region and with no strong absorption bands \citep{Fraeman_2012, Poggiali_2022}. This may be due to several reasons: (1) The surface of Phobos may be desiccated and strongly altered by space weathering \citep{Poggiali_2022}. (2) The presence of opaque materials could severely reduce the absorption bands \citep{Cloutis_1990c, Cloutis_2009}. (3) The size of the grains, which has an impact on the absorption bands, could also prevent their detection as absorption features are shallower for fine grains \citep{Sultana_2021}. On the other side, organic compounds have been observed on dark and primitive asteroids, especially for C-, B-, and P-types which can exhibit a 3.4 µm absorption band \citep{Hromakina_2022}, as well as on D-type like Jupiter Trojans \citep{Brown_2016, Wong_2023}. \par
In this context, D-type asteoroids, also thought to be very primitive and with a spectral slope close to Phobos spectrum, are particularly rare and seldom observed. They are usually found in the main belt or in the Trojans and known to be spectrally close to comets \citep{Fornasier_2007, Vernazza_2017}. For the 67P/Churyumov-Gerasimenko comet, a large part of the comet's mass could be associated with organic compounds \citep{Quirico_2016, Bardyn_2017, Choukroun_2020}. The Rosetta mission has observed infrared signatures of organics, especially near 3.4 µm due to aliphatic carbon \citep{Raponi_2020}. Hence, if Phobos comes from a captured asteroid, it could be expected that organic materials are also found on Phobos. \par
This range of the spectrum will be measured on Phobos for the first time using the MIRS spectrometer \citep{Barucci_2021} onboard MMX. MIRS is an infrared spectrometer (0.9 - 3.6 µm) developed by the LESIA laboratory with the support of CNES and several institutions in France and Japan. It will measure Phobos spectra with high spatial and spectral resolution (resp. up to 20 m/px and $\sim$20 nm). The MMX mission will also bring back samples (> 10 g) from Phobos to Earth in 2031 which will allow precise measurements of the organic content of Phobos. \par
This part of this study is a continuation of the work presented in \cite{Wargnier_2023_1}. We used here different organic compounds (with aliphatic/aromatic signatures) compared to the previous study, which is more representative of the type of organics expected on Phobos in the case of the captured asteroid hypothesis. We searched to assess the detectability of organic materials (i.e. the necessary quantity of organics to detect the bands) in a Phobos spectroscopic simulant by adding different volume fractions of CH-rich organics. We are also interested in the detection level of phyllosilicates in the simulants. The presence of hydrated minerals has already been discussed for Phobos (e.g., \citealt{Giuranna_2011, Fraeman_2014, Glotch_2018}). Similarly to organics, the hydroxyl feature at 2.7 µm has been observed on numerous primordial asteroids \citep{Takir_2012, Rivkin_2022} and recently detected on (162173) Bennu \citep{Hamilton_2019} and (101955) Ryugu \citep{Kitazato_2019} by OSIRIS-REx and Hayabusa 2 observations as well as on their samples \citep{Pilorget_2022, Nakamura_2023, Hamilton_2024}. Phyllosilicates are expected to be detected in the case of the captured asteroid hypothesis. However they may also be detected in the giant impact scenario with a low heating formation process and high water content \citep{Ronnet_2016, Pignatale_2018}. It is worth noticing that no unambiguous indication for this usually deep band was detected by varied observations of Phobos. Ground-based observations are hampered, in the 2.7 µm region, by the atmosphere but can still be consistent with the presence of a hydroxyl feature \citep{Rivkin_2002, Takir_2022}. Based on CRISM observations, \cite{Fraeman_2014} found the possible presence of asymetric band edges at 2.7 µm although the CRISM data between 2.70 and 2.76 µm is not usable.\par
In the third objective of our work, we search to improve the spectral match of Phobos to our simulants \citep{Wargnier_2023_1, Wargnier_2023_2} in the mid-infrared range. We create a new Phobos simulant adding phyllosilicates into the initial mixture (olivine, anthracite, DECS-19 coal, see \citealt{Wargnier_2023_2}) as suggested by \cite{Giuranna_2011}. We prepared varied samples of powdered materials to reproduce characteristic features of the Phobos spectrum from the VNIR to the MIR. We noticed however that other works propose materials and mixtures as simulants to Phobos composition \citep{Pajola_2013, Rickman_2016, Landsman_2021, Miyamoto_2021}. Two of the most used in the community are the Tagish Lake meteorite and the University of Tokyo Phobos Simulant (UTPS, \citealt{Miyamoto_2021}). Therefore, we compare results on our simulants with the results obtained on the Tagish-based UTPS (UTPS-TB) simulant.\par
We also investigate the effect of the geometry of observation which is an important parameter as it affects the slope and reflectance. Specifically, the phase curve, which represents the evolution of reflectance with phase angle, is critical to understand the texture of the surface in addition to spectroscopic observations, more sensitive to the composition. \par
Finally, we remind the readers that this paper is the first in a series on the spectro-photometry of Phobos simulants. We plan studying the spectro-photometric modifications of different Phobos simulants adding the effect of several properties of a surface and processes that can alter the surface of a material (e.g., space weathering, increase of porosity/roughness,...). Therefore, we will be able to compare the results with the observations of Phobos taken at different phase angles from ground-based and remote-sensing observations with laboratory data, and help to constrain the Phobos surface properties from the future MIRS observations.

\section{Materials and methods}

\subsection{Preparation of samples, mixtures, and simulants}
This work presents the results of three main investigations: (1) the evaluation of the 2.7 µm and 3.4 µm features in Phobos-like material, (2) the update of our previous spectroscopic simulant \citep{Wargnier_2023_2}, and (3) the computation of photometric properties of Phobos simulants.
To achieve these results, we used various materials including silicates, phyllosilicates, organics, and opaque materials. The endmembers were selected based on their representativeness in the context of the Phobos surface, as well as their spectral properties that match well with the Phobos spectra in the visible, near-infrared, and/or mid-infrared. Some of these endmembers were previously presented in other works \citep{Wargnier_2023_1, Wargnier_2023_2}. We have previously discussed the presence of olivine and opaque materials. The olivine used in this work was purchased from Donghai Crystal products and is a green forsterite with half-centimeter-sized crystals.  It has a very similar Mg/Fe ratio (Sect. \ref{sect:sample_characterization}) and spectral properties (Sect. \ref{sect:results_spec}) compared to the previous olivine. The spectra of Phobos in the visible and near-infrared (VNIR) range can be accurately replicated using dark opaque materials. To achieve a flat, dark, and featureless spectrum in the VNIR range, darkening agents are particularly useful. In literature, several opaque materials were found (i.e., FeS and Fe$_{3}$O$_{4}$) to be common minerals in carbonaceous chondrites. Despite the good mineralogical representativeness of FeS, we found that this material reproduces poorly the red spectral slope of Phobos in the near-infrared \citep{Wargnier_2023_1}. Therefore, in this study, we selected anthracite as darkening agent due to its optical properties and the ability to reduce absorption bands and the reflectance level \citep{Sultana_2023}. The reader can refer to \cite{Wargnier_2023_1} for more details about anthracite. Despite its interesting optical properties, anthracite is not a plausible analog material for the surface of a small body, as its formation requires a high pressure and high temperature ($\sim$200$\degree$C) over a period of more than 0.1 Myr \citep{Quirico_2016}. This is a very mature coal with about 92\% of carbon content. The use of anthracite in this study is solely for the purposes of simulating opaque materials and not as a representative organic material for the Phobos surface. Furthermore, the utilisation of anthracite as opaque material was primarily driven by its ease of use in the context of a simulant, in comparison with FeS, which tends to rapidly oxidise. Elementary composition, grain size, density, and origins are presented for each endmembers in Table \ref{table:endmembers}. \par
Several papers have discussed the presence of phyllosilicates on the surface of Phobos \citep{Giuranna_2011, Fraeman_2012, Fraeman_2014, Glotch_2018}. The presence of hydrated minerals can provide valuable insights into its origins. The mid-infrared (MIR) spectra of Phobos were analyzed using data from the Mars Express Planetary Fourier Spectrometer (PFS) and the Thermal Emission Spectrometer (TES) onboard Mars Global Surveyor. The results indicate that a linear mixture of antigorite and biotite phyllosilicates can well reproduce the MIR spectrum of Phobos \citep{Giuranna_2011, Glotch_2018}. Phobos spectra show also a good agreement with a nepheline spectrum \citep{Giuranna_2011}. Therefore, we included antigorite and biotite, two phyllosilicates in the Phobos simulant and also nepheline to experimentally match the MIR Phobos spectrum. Antigorite is a phyllosilicate from the serpentine subgroup. It was initially a green-brown piece of about one centimeter and 1.1 grams kindly provided by the Muséum National d'Histoire Naturelle (MNHN) of Paris. Biotite, a bright black mineral from Canada, is a phyllosilicate from the micas family. The sample was purchased from Kremer Pigmente (\#53220.12100.136) as a fined-grain powder (0-250 µm). Nepheline, specifically eleolite, is a greyish-brown feldspathoid mineral that is rich in aluminium. It was also provided by the MNHN in the form of three small pieces of about one gram in total. Its origin is from Brevig in Norway. Although these three endmembers were detected (as minor phases for biotite and nepheline) in some carbonaceous chondrites \citep{Rubin_1997} their presence in Phobos composition is not unambiguously expected in the giant impact formation scenario \citep{Sautter_2021, Cuadros_2022, Liu_2023}. Therefore, we added to our sample set also saponite \citep{McLennan_2014} as a primitive martian crust material. However, interestingly, saponite was also detected as a major component in Ryugu grains (e.g., \citealt{Nakamura_2023, Nakato_2023}). Saponite SapCa-2 (California, USA) was purchased from the Clay Mineral Society. It is a greenish/brownish mineral of the smectite family.\newline
The presence of organics was already discussed for Phobos in the Introduction (Sect. \ref{sect:intro}), and it is only expected in the case of the captured asteroid hypothesis. In this scenario a coal from the Penn State Coal Sample Bank (DECS-19) is particularly useful for its strong red slope in the visible \citep{Wargnier_2023_2}. DECS-19 is a low-volatile bituminous coal provided by the Coal Sample Data Bank from Penn State University. This mature coal has a mean-maximum vitrinite reflectance of 1.71\%, H/C = 0.65, O/C = 0.027 and N/C = 0.01. Ashes account for $\sim$4.6\%, and total sulfur form for 0.74\%. Coals have been used as optical analogs of extraterrestrial insoluble organic matter (IOM, \citealt{Moroz_1998}), though their structure and composition display substantial differences \citep{Quirico_2009, Quirico_2016, Alexander_2017}. DECS-19 is a quite mature coal, which better fits with organic matter hosted by a thermally metamorphosed chondrite, and possibly with organics in micro-meteorites that were heated through collision with Phobos, thereby generating its regolithic surface. Titan tholins were used to reproduce the simulant proposed in \cite{Wargnier_2023_1}. The preparation of the different materials was similar to \cite{Wargnier_2023_1}: endmembers were ground using a cryogenic grinder Retsch Cryomill at Laboratoire Génie des Procédés et Matériaux (LGPM). A single steel ball with a diameter of one centimetre was used for dry grinding. Endmembers were then successively dry-sieved in different grain size ranges: 0-25 µm, 25-50 µm, 50-100 µm, and 100-450 µm. Because we expect for Phobos grains a size of about 80 µm \citep{Kuehrt_1992, Giuranna_2011, Miyamoto_2021} we chose, in particular, to extract grains between 25-50 µm and grains between 50-100 µm. Biotite was purchased pre-grounded and then sieved using the same protocol. Hyperfine dark materials were already prepared as submicrometric particles. The protocol to obtain the hyperfine dark powders is described in \cite{Sultana_2021} and \cite{Sultana_2023}. This increases the darkening power of these materials. Although extraterrestrial materials do not necessarily exhibit such a distribution of opaque material sizes, the Ryugu samples demonstrate the presence of very fine opaque grains in relatively large quantities (e.g., \citealt{Nakato_2023}). After, the endmembers were mixed together using an agate mortar to obtain an intimate mixture. Mixtures were made in volume percentages \citep{Wargnier_2023_2}. Density values for some endmembers have been obtained directly for our samples. This is the case for olivine and anthracite \citep{sultana2021_these}. Other mean density values were obtained from the literature. Prepared mixtures for this work are described in Table \ref{tab:composition_mixtures}. \par
Some mixtures were also already prepared for other studies and used directly here or reproduced. These mixtures will be referred to as "Phobos simulants" throughout the text. The University of Tokyo Phobos Simulant (UTPS, \citealt{Miyamoto_2021}) was the first simulant of Phobos that we used. UTPS is a Tagish Lake-based simulant that roughly reproduces the main Phobos spectral properties. It was obtained from the University of Tokyo as a 500-gram piece. Regarding the pure minerals, the UTPS was ground and sequentially sieved to obtain simulant grains of different sizes. The second simulant is described in \cite{Wargnier_2023_1}. This simulant is composed of olivine, anthracite, and tholins and is used for detectability purposes. Its spectral properties in the VNIR, including slope and reflectance, are well-matching the characteristic of Phobos simulant although the composition is not particularly close to the Martian moon. Additionally, the simulant has the advantage of showing no absorption in the relevant wavelength range, making it an excellent choice for detectability studies on a Phobos-like surface.

\begin{figure*}
\resizebox{\hsize}{!}{\includegraphics{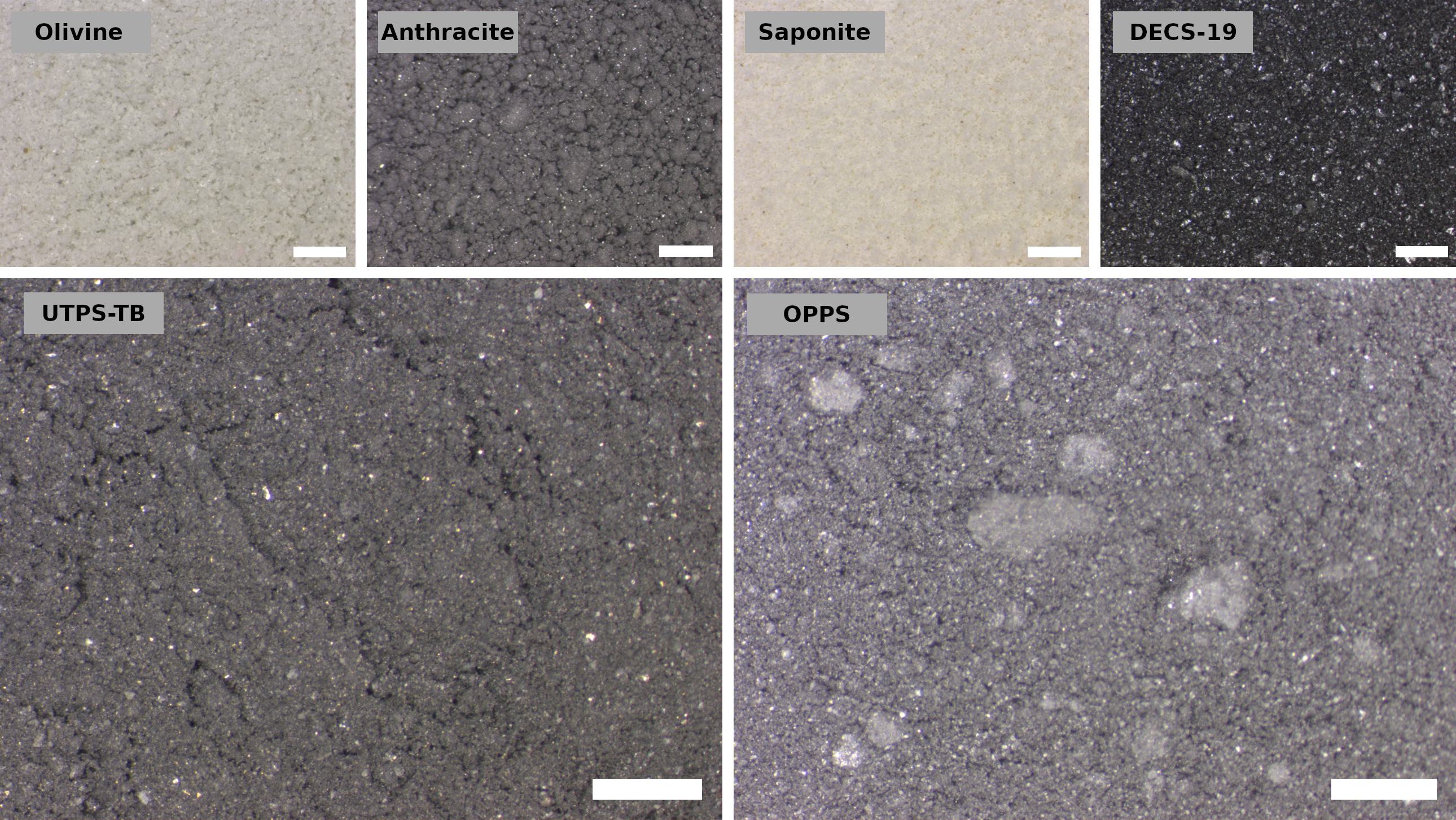}}
\caption{Mosaic of optical images of the surfaces of the two Phobos simulants (UTPS and OPPS) that were investigated, as well as the endmembers of the OPPS. The bright spots visible in OPPS correspond to those in DECS-19. DECS-19 grains tend to have large facets and diffuse light in specular reflection, resulting in bright spots. Each scale bar represents 1 mm. Grains morphology and size were investigated using SEM images (see Fig. \ref{fig:SEM_images}).} 
\label{fig:opt_micro_image}
\end{figure*}

\subsection{Samples characterization} \label{sect:sample_characterization}

Elemental investigation of endmembers was obtained using energy-dispersive X-ray spectroscopy (EDX) and for the biotite sample, Laser-Induced Breakdown Spectroscopy (LIBS). EDX analysis was performed using an EDAX Metek New XL30 coupled to a FEI Quanta200 environmental SEM at LGPM with an accelerating voltage of 15 kV. LIBS emission spectrum of biotite was obtained using the TX1000 instrument from Iumtek. We performed SEM images in low vacuum mode with magnification between 250 and 4000.\par
EDX analysis reveals an Mg-rich composition for the olivine (Mg\# defined as the atomic ratio Mg/(Mg+Fe) = 0.90). We performed also EDX on the various phyllosilicate samples. For example, antigorite appears to be mostly dominated by magnesium. It is also composed of silicon and iron. EDX analysis of these endmembers and of the phyllosilicates is presented in Table \ref{table:EDX}. LIBS spectrum has shown that biotite is mostly composed of aluminum, magnesium, iron, silicon, potassium, and calcium. We observed also a minority of lithium, sodium, and titanium; and traces of rubidium, strontium, beryllium, barium, chromium, zinc, and tin. For the three last elements, the content is about 20 ppm. Light elements such as oxygen, nitrogen, and hydrogen cannot be investigated with the LIBS instrument. Chromium, zinc, and tin are not expected to be found in biotite but could be due to the grinding process. Lithium, sodium, and titanium are classically found as traces in biotite samples \citep{Ellis_2022, Redin_2023}. The LIBS instrument is here particularly sensitive to these two first elements. Whereas calcium is found as a major element in the biotite, it is not expected. However, it has already been observed in biotite samples and was explained by contamination of the natural biotite by calcic rocks \citep{Lovering_1972}. LIBS spectrum of biotite in \cite{McMillan_2007} shows also important calcium emission bands.\par
DECS-19 was already fully characterized and additional information can be found on the Penn State Coal Sample Bank website. Saponite SapCa-2 was characterized by the Clay Mineral Society and information can be found on their website. We also reported elemental composition of phyllosilicates given by the EDX measurements in Appendix (Table \ref{table:EDX}).\par
For grain size measurements, we used both SEM images -- which allow also to analyze the state of the mixture -- and a laser diffraction analyzer. The particle size analyzer is a Malvern Mastersizer 3000 with the Hydro MV accessory. We dispersed the samples in ultrapure water (15 MOhm.cm) using both stirrer and sonicator to disrupt the aggregates that may form. We acquired data for 30 seconds with two laser at different wavelength to cover different grain size ranges and repeat at least 6 times the measurements. Averaging the measurements, we can obtain a particle size distribution of our samples in a wide range, from nanometers to millimeters. For UTPS and olivine we obtained a distribution (Fig. \ref{fig:granulo} in Appendix) peaked at 50-100 µm, as expected because the samples were sieved at this grain size range. Although most of the distribution is in this range, we also noticed a tail towards small grains. This is because small grains tend to clump together with larger grains, forming relatively large aggregates that do not pass through the sieve, but are disrupt by the sample preparation for the particle size analyzer. For the anthracite, the laser granulometry was probably less efficient because the SEM images showed that most of the grains are hyperfine ($<$1 µm), but the laser method shows a complicated behaviour of the distribution with several modes. In particular, even with the precaution of disrupting the aggregate, we noticed that some grains re-aggregate after some time. Laser diffraction granulometry is then a useful complementary method to SEM and the combination of the two techniques allows the full characterisation of the grain size of the samples. 

\begin{table*}
	\centering
	\caption{Composition of the prepared mixtures in olivine, anthracite, DECS-19, and phyllosilicates. Quantities are given in vol.\%. PSD \citep{Wargnier_2023_1} is composed of olivine (77 vol.\%, 100 µm), anthracite (20 vol.\%, $<$1 µm), and Titan tholins (3 vol.\%, 400 nm). Tholins was previously used in \cite{Wargnier_2023_1} to obtain a red slope, close to the Phobos' spectral slope. PSD (Phobos simulant for detectability) is only used for the detectability study as it exhibits no absorption associated to a reflectance and spectral slope similar to Phobos spectra. Mixing ratios for the simulant mixtures were determined after several attempts based on their spectroscopic properties \citep{Wargnier_2023_1, Wargnier_2023_2}. Name of the simulant mixtures were determined by the phyllosilicate composition: ATG for a simulant containing antigorite, BIO for biotite, NEP for nepheline, and SAP for saponite. Each of these mixtures contains also olivine, anthracite, and DECS-19 \citep{Wargnier_2023_2}.}
	\label{tab:composition_mixtures}
    \resizebox{\textwidth}{!}{
	\begin{tabular}{cccccccccc} 
		\hline
		&  & \textbf{PSD} & \textbf{Olivine} & \textbf{Anthracite} & \textbf{DECS-19} & \textbf{Antigorite} & \textbf{Biotite} & \textbf{Nepheline} & \textbf{Saponite}\\
        & Grain sizes &  & 50-100 µm & $<$1 µm & 50-100 µm & 50-100 µm & 50-100 µm & 50-100 µm & 50-100 µm\\
        & \textbf{Mixtures} &  & & & & & & & \\
		\hline
		  \parbox[t]{2mm}{\multirow{11}{*}{\rotatebox[origin=c]{90}{\textbf{Detectability tests}}}}  & CH-rich-5 & 95 & -- & -- & 5 & -- & -- & -- & --\\
		  & CH-rich-10 & 90 & -- & -- & 10 & -- & -- & -- & --\\
		  & CH-rich-15 & 85 & -- & -- & 15 & -- & -- & -- & --\\
		  & CH-rich-20 & 80 & -- & -- & 20 & -- & -- & -- & --\\
		  & CH-rich-30 & 70 & -- & -- & 30 & -- & -- & -- & --\\
		  & hydmin-rich-1 & 99 & -- & -- & -- & 1 & -- & -- & --\\
		  & hydmin-rich-3 & 97 & -- & -- & -- & 3 & -- & -- & --\\
		  & hydmin-rich-5 & 95 & -- & -- & -- & 5 & -- & -- & --\\
		  & hydmin-rich-10 & 90 & -- & -- & -- & 10 & -- & -- & --\\
		  & hydmin-rich-15 & 85 & -- & -- & -- & 15 & -- & -- & --\\
          & hydmin-rich-20 & 80 & -- & -- & -- & 20 & -- & -- & --\\
        \hline
           \parbox[t]{2mm}{\multirow{11}{*}{\rotatebox[origin=c]{90}{\textbf{Simulant mix.}}}} & 
              SIM-ATG-BIO-1 & -- & 30 & 20 & -- & 25 & 25 & -- & --\\
            & SIM-ATG-BIO-2 & -- & 25 & 15 & 10 & 25 & 25 & -- & --\\
            & SIM-ATG-BIO-3 & -- & 20 & 20 & 20 & 20 & 20 & -- & --\\
            & SIM-ATG-BIO-4 & -- & 15 & 25 & 10 & 25 & 25 & -- & --\\
    		& SIM-NEP-1 & -- & 25 & 15 & 10 & -- & -- & 50 & --\\
            & SIM-NEP-2 & -- & 20 & 20 & 20 & -- & -- & 40 & --\\
            & SIM-NEP-3 & -- & 15 & 25 & 20 & -- & -- & 40 & --\\
            & SIM-ATG-SAP-1 & -- & 25 & 15 & 10 & 25 & -- & -- & 25\\
    		& SIM-ATG-SAP-2 & -- & 20 & 20 & 20 & 20 & -- & -- & 20\\
            & SIM-SAP-1 (OPPS) & -- & 20 & 20 & 20 & -- & -- & -- & 40\\
		\hline
	\end{tabular}
    }
\end{table*}

\subsection{Reflectance measurements}
In this study, we performed bidirectional reflectance spectra from visible to mid-infrared using different instruments. Reflectance spectra in the VNIR (0.6 - 4.2 µm) were made at room temperatures using the Spectro-photometer with cHanging Angles for Detection Of Weak Signals (SHADOWS, \citealt{Potin_2018}) at Institut de Planétologie et d'Astrophysique de Grenoble (IPAG, France). We acquired also spectra in the VNIR with the SpectropHotometer with variable INcidence and Emergence (SHINE, \citealt{Brissaud_2004}), also hosted at IPAG. SHINE is a VNIR spectrophoto-goniometer that measures reflectance spectra from 0.6 to 4.2 µm. SHINE was used in the Gognito mode \citep{Potin_2018} which allow us to reduce the size of the illumination, and hence to observe smaller and darker samples. These spectrogonio-radiometers measure in both visible and near-infrared using two different detectors: a silicon detector for the visible part and an InSb detector cooled at 77K for the NIR. The spectral resolution of SHADOWS and SHINE is wavelength-dependent, spectral sampling is 20 nm. Spectral resolution of the spectrogognio-radiometers is comparable to the expected MIRS spectral resolution \citep{Barucci_2021}. For each wavelength, the intensity is obtained with a 300 ms integration time. This measurement is repeated 20 times. The calibrated reflectance is then given by the mean value of these 20 measurements for a wavelength and the error on the measurement is given by the standard deviation. Instrumental and atmospheric effects were suppressed by taking reference measurements using a Spectralon\textcopyright{} reference in the visible and an Infragold\textcopyright{} reference in the near-infrared. In this paper, unless otherwise stated, we used the nominal SHADOWS and SHINE geometrical configuration (incidence i = $0\degree$, emission e = $30\degree$, and azimuth $\phi$ = $0\degree$). Measurements at different phase angles have been also performed from 5$\degree$ to 130$\degree$. Due to instrument limitations, spectra at phase angle smaller than 5$\degree$ cannot be obtained.\newline
Near- and mid-infrared spectra were obtained with a Fourier-Transform Infrared Spectrometer (FTIR) Bruker VERTEX 70V (1.25 - 20 µm) at IPAG. The instrument is mounted with a Bruker accessory (A513/QA) for bidirectional measurements. The MCT detector is cooled with liquid nitrogen (LN-MCT) and the sample chamber is purged with dry air. For each measurement, we performed 100 scans with a spectral resolution of 4 cm$^{-1}$. Reference was taken with direct flux with i = $90\degree$ and e = $90\degree$. Using this method for reference, the spectrum is not in absolute unit. Therefore, a calibration procedure has been defined for FTIR spectra using a multiplicative factor to overlap the spectrum measured with SHADOWS. To take into account the possible variation of the sample surface, we measured the reflectance spectrum for each sample three times, re-preparing each time the sample. 

\subsection{Spectral and photometric analysis}
\subsubsection{Reflectance, slope, and band depth} \label{sect:bd_compute}
In the VNIR, three main parameters were studied: the spectral slope, the reflectance and the band depth. Spectral slopes were obtained 720 and 900 nm, 1.5 µm and 2.4 µm, and 3.7 and 4.0 µm using the formula given in \cite{Wargnier_2023_1}. The spectra were normalized and the slope then computed at the starting wavelength of each interval, namely 720 nm, 1.5 µm, and 3.7 µm.\par
\begin{figure}
\centering
\resizebox{8cm}{!}{\includegraphics[width=\textwidth]{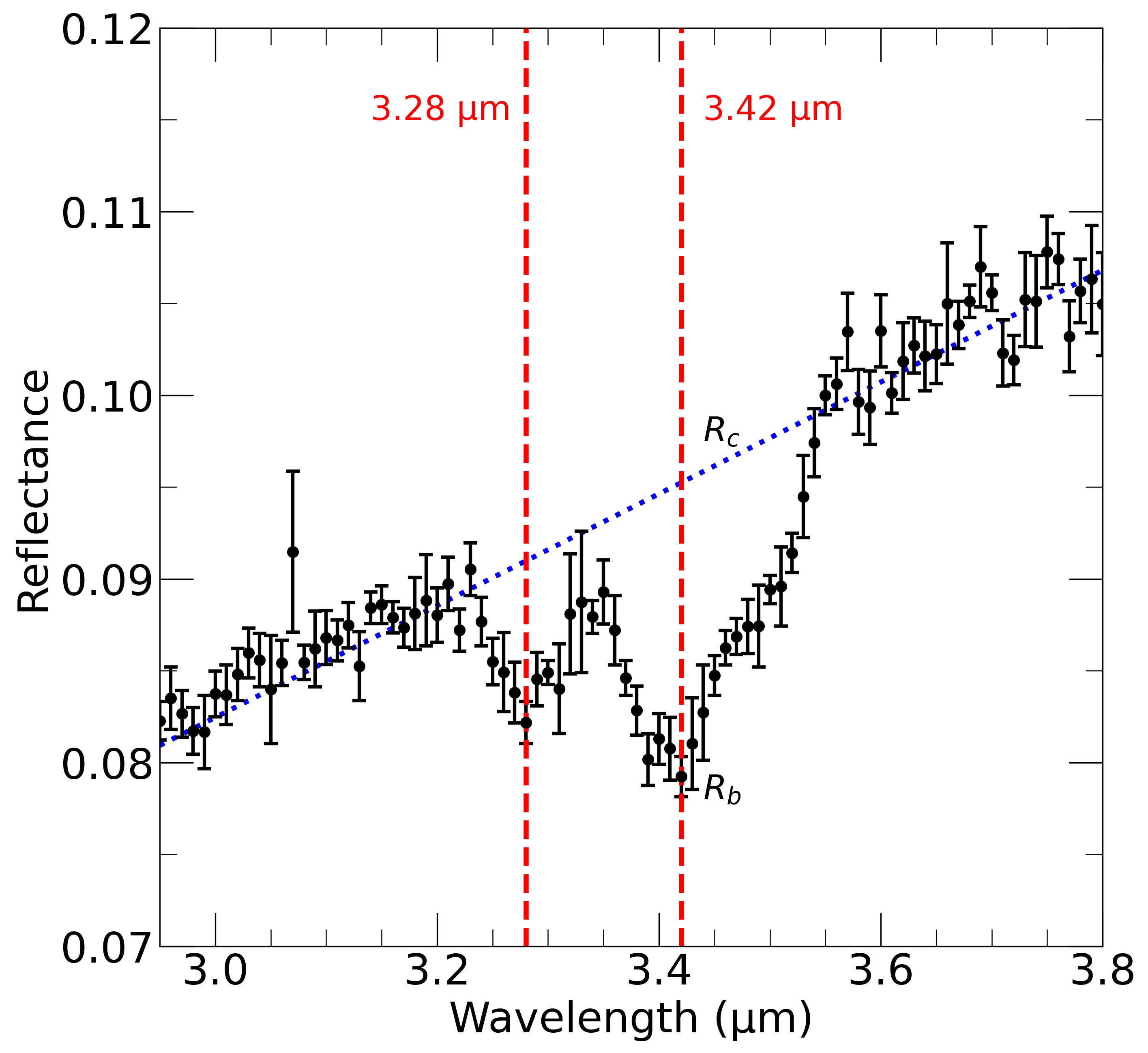}}
\caption{SHADOWS spectrum with the associated uncertainties of the 3.4 µm region of the mixture CH-rich-30. Uncertainties were measured by SHADOWS with the standard deviation of ten points for a given wavelength. Red dashed vertical lines represent the 3.28 µm and 3.42 µm wavelength. Blue dotted line is the continuum. Reflectance of the continuum at 3.42 µm is noted $R_c$ and corresponds to the reflectance of the continuum. $R_b$ is defined as the reflectance in the band. By obtaining these parameters we can determine the band depth.} 
\label{fig:method_BD}
\end{figure}

Reflectance was computed at 600 nm, 1.8 µm, and 4.0 µm directly from the original spectrum. Bands depth were computed using the following procedure: the absorption bands at 3.28 and 3.42 µm were characterized using a linear fit for the continuum as shown in Fig. \ref{fig:method_BD}. Anchor points for the linear fit were fixed at different areas near the edge of the features for each sample. This allows obtaining an uncertainty on the definition of the continuum. We take also into account the uncertainties of the SHADOWS measurements (Fig. \ref{fig:method_BD}). We chose to not include the 3.0 µm absorption band in the linear fit because we focused our work on the 3.4 µm and because this feature is mainly due to water absorption in our sample. However, due to the presence of the 3.0 µm feature, the 3.42 µm absorption band will be probably a bit overestimated compared to the 3.28 µm absorption feature. The band depth was computed using the formula from \cite{Clark_1984}. X-axis error bars have been obtained with the uncertainty of the balance. Y-axis error bars have been computed using the propagation of uncertainties and by taking into account the two sources of uncertainties described above. The hydrated mineral band depth was computed using a similar method, with anchor points for the continuum linear fit defined at 2.4 and 4.0 µm.  

For MIR spectra, we measured reflectance spectra and converted spectra in emissivity using the Kirchoff's law: $E = 1 - R$, where R is the reflectance and E the emissivity \citep{Salisbury_1991, Salisbury_1994}. Although it is an approximation of an emissivity spectrum, it reproduced well the position of bands, and is useful for qualitative study and comparison with in situ MIR spectra of Phobos \citep{Sultana_2023}. Position and shapes of bands are not significatively modified \citep{Salisbury_1991, Martin_2024}. In particular, we focused on phyllosilicates bands, Restrahlen bands (RB), Transparency features (TF), and Christiansen features (CF). These three last features are particularly important for compositional identification. CF generally corresponds to a maximum of emissivity and is caused by an important change in the refractive index of the material (e.g., \citealt{Salisbury_1991, Martin_2022}). RBs are due to fundamental molecular vibration bands \citep{Salisbury_1991} whereas TF depends on the low absorption coefficients in some wavelength region due to the presence of grain with size smaller than 75 µm \citep{Salisbury_1989,Salisbury_1991, Hamilton_2000, Morlok_2019, Martin_2022}. Hence, TF depends on the composition of the materials \citep{Salisbury_1991}. TF can be generally observed around 10-12 µm (e.g., \citealt{Mustard_1997, Mustard_2019, Pisello_2022, Poggiali_2023, Morlok_2023}). \par
In this work, we compared experimental spectra with the TES red unit spectrum of Phobos \citep{Glotch_2018}. Main features observed in this spectrum are: a double CF at 8.49 µm and 9.17 µm, a RB at 9.8 µm, and a TF at 12.4 µm.

\subsubsection{Hapke modeling} \label{sect:hapke_model}
To compare our results with other Phobos observations or with other bodies and laboratory experiments, it is particularly interesting to apply the Hapke model as explained below. 
\paragraph{The model} \mbox{} \\
For this work, we used the Hapke IMSA model \citep{Hapke_2012a} for bidirectional reflectance with a two term single-scattering phase function:

\begin{align}
r(i,e,\alpha) &=\frac{K \omega}{4} \frac{\mu_{i}}{\mu_{i}+\mu_e} S(i,e,\alpha,\bar{\theta}) \nonumber \\                  
              & \times \left[P_{hg}(\alpha, b, c) \cdot B_{sh}(\alpha, B_{sh,0}, h_{sh}) \right. \nonumber\\ 
              & \left. + B_{cb}(\alpha, B_{cb,0}, h_{cb}) \cdot M \left(\frac{\mu_i}{K}, \frac{\mu_e}{K}, \omega \right) \vphantom{\frac{\mu}{K}} \right] 
\label{hapke_global}
\end{align}

The model has an important number of parameters and functions: $\alpha$ is the phase angle, $\omega$ is the single scattering albedo, $\mu_{i}$ and $\mu_{e}$ are respectively equal to cos(i) and cos(e) -- where i is the incidence angle and e the emission angle. \newline
The B$_{sh}$ and B$_{cb}$ function describes the opposition effect. Each of these functions take into account for different physical effect that occurs at small phase angle: the coherent backscattering effect (CBOE) and the shadow-hiding opposition effect (SHOE). The reader is referred to \cite{Hapke_2002} for more details on these effects. The CBOE occurs for very small phase angle ($<$ 1$\degree$, \citealt{Helfensetein_1997, Hapke_1998}) and considering the lack of data at these phases due to instrument limitations, the contribution of this function can be ignored. The SHOE function is defined as:
\begin{equation}
    B_{sh}(\alpha, B_{sh,0}, h_{sh})=\frac{B_{sh,0}}{1+\frac{\tan \alpha/2}{h_{sh}}}
\end{equation}
where h$_{sh}$ is the half-width of the SHOE. In principle, $B_{sh,0}$ is expected to be between 0 and 1 \citep{Hapke_1986}. However, new models have shown that values of the opposition effect intensity up to 3 can be observed for the rough and irregular surfaces observed on small bodies \citep{Li_2015}. \newline
The K parameter has been added in \cite{Hapke_2008} and correspond to the porosity factor. We will use here the approximation of the porosity factor given in \cite{Helfenstein_2011}: 
\begin{equation}
    K \approx 1.069 + 2.109 h_{sh} + 0.577 h_{sh}^2 - 0.062 h_{sh}^3
\end{equation}

$P_{hg}$ is the double-term Henyey-Greenstein (2T-HG) function \citep{McGuire_1995}: 
\begin{align}
    P_{hg}(\alpha, b, c) &= \frac{1+c}{2}\frac{1-b^2}{(1+2b \cos \alpha + b^2)^{3/2}} \nonumber \\ 
    & + \frac{1-c}{2}\frac{1-b^2}{(1-2b \cos \alpha + b^2)^{3/2}}, 
\end{align}
where b and c are related to the scattering behavior of a particle: b is linked to the phase function shape and c to the type of scattering. A negative c value indicates forward scattering whereas a positive c corresponds to backscattering. It is important to note that there are multiple definitions of 2T-HG in the literature, which can result in slightly different bounds for the c parameter. M is the Hapke's multiple scattering function \citep{Hapke_2002} and the macroscopic roughness function S is presented in details in \cite{Hapke_1993}. One of the parameters of this function is the surface roughness $\bar{\theta}$ \citep{Hapke_1984}. This parameter dominates for large phase angle data \citep{Helfenstein_1989}. \newline
We searched for the Hapke parameters: $\omega$, b, c, B$_0$, h, and $\bar{\theta}$. These six free parameters give physical information of the surface \citep{Hapke_1981, Hapke_2002}. We searched the parameters inside the following boundaries: w = [0,1], b = [0,1], c = [-1,1], B$_{0}$ = [0,3], h = [0,0.15], $\theta$ = [11,50]. The boundaries were chosen to explore parameters within the typical range of laboratory experiments (\citealt{Beck_2012, Pommerol_2013}) and to include photometric parameters determined for Phobos \citep{Fornasier_2023}. The model is applied on phase curve obtain using reflectance spectra at different geometries of observation. We followed the recommendation from \cite{Schmidt_2015} and applied the Hapke modeling on the entire dataset at the same time, including measurements at different incidence angle.
\paragraph{Inversion methods} \mbox{} \\
The Hapke model inversion was performed using various methods because the Hapke equation is non-linear equation with highly correlated parameters. Therefore, the equation generally shows several local minima which makes it difficult to converge to the global minimum. For all these methods, we tried to minimized the reduced $\chi^{2}$: 
\begin{equation}
    \chi^{2} = \sum_{i=1}^{n} \frac{\left(r_{i, exp} - r_{i, Hapke}(i,e,\alpha)\right)^2}{\sigma_{i}}
\end{equation}
Initially, we attempted to use a Levenberg-Marquardt (LM) algorithm, which is based on the gradient and Gauss-Newton algorithms. However, the LM algorithm is more stable than the Gauss-Newton alone. Nonetheless, a good estimate of the initial Hapke parameters is still required. The LM routine was then run with various and random initial parameters (not too far from the estimate). This method enables us to determine whether we converge to local minima instead of the global minimum.\newline
We also tried to fit our experimental using a basin-hopping algorithm. Basin-hopping is a global optimization algorithm that performs several cycles of local optimization combined to random perturbation/step around the solution. At each step, a Metropolis test is performed. We chose to use a Broyden-Fletcher-Goldfarb-Shanno (BFGS) algorithm as local optimizer. BFGS is a quasi-Newton method mainly based on approximation of the Hessian matrix to solve nonlinear optimization problem.\newline
We finally tested a Bayesian method to fit the Hapke model using Markov Chain Monte Carlo (MCMC) methods based on the \textit{emcee} MCMC python implementation \citep{Foreman_2013}. This method takes longer to run, but one of its advantages is that it provides a a-posteriori probability density function (PDF) of the values of the different parameters. This is particularly useful because nonlinear problem (such as Hapke's equation inversion) with several parameters does not have a unique solution and bayesian inference allows to explore the parameter space. For the six Hapke parameters, we considered a prior uniform PDF. We performed the MCMC for 50000 steps and considering a burn-in period of 5000 steps. The best parameters were chosen from the median of the posterior probability distribution. We also computed the Maximum Likelihood Estimation (MLE) in order to obtain more information about the results. With this setup for the MCMC inversion, we obtained a typical mean acceptance fraction of 0.283. \par
To give an indication about the difference between the data and the fit, the results of the different methods will be presented associated with the the residuals and/or with the root mean square residual (RMS). Considering the methods explored in this work, it appears that MCMC is a very good approach to inverse the non-linear Hapke equation with a relatively high number of free parameters. However, the bayesian inference is a very computation time-consuming method and while the number of point in our dataset is quite small, it can be more complicated to used this Monte Carlo strategy for large dataset. 
\paragraph{Parameters uncertainties} \mbox{} \\
Determining uncertainties in the inversion of the Hapke model can be challenging. They are often under- or over-estimated because all the parameters are correlated. Depending on the methods, we estimated the uncertainties with various manners: (1) For the LM algorithm we took the uncertainties as the diagonal terms of the covariance matrix. However, the uncertainties appears so large that we decided to computed the uncertainties based on the variability based on the fit when varying the initial conditions and parameters space. This method appears unfortunately also prone to large errors. (2) For the basinhopping method, we obtained the uncertainties with the inverse of the Hessian matrix, and (3) For the MCMC inversion, we assumed that the uncertainties are given by the 16-50-84 percentiles. \par 
Several others studies were trying to have a good estimate of the uncertainties when fitting the Hapke model (e.g., \citealt{Gunderson_2006, Shepard_2007, Schmidt_2015}). \cite{Shepard_2007} estimated the uncertainties by taking into account the variation of the obtained parameters values when fitting variants description of the Hapke model. \cite{Schmidt_2015} used synthetic datasets and inversion using MCMC to explore the parameters space and obtained realistic uncertainties for Hapke modeling of photometric measurements. They found that uncertainties on the Hapke parameters can be less than 10\% in very favorable conditions such as with a full BRDF from almost 0 to 180 degrees of phase angle, and/or with small uncertainties on the data (< 5 \%), etc. Therefore, there is not a unique method to obtain uncertainties and the coverage of the dataset in terms of illumination angles can drastically change the values of the parameters and especially of the associated uncertainties. 

\section{Results}
\subsection{Spectral properties of the endmembers} \label{sect:results_spec}
\subsubsection{In the VNIR}
\begin{figure*}
     \centering
     \begin{subfigure}[b]{0.49\textwidth}
         \centering
         \resizebox{\hsize}{!}{\includegraphics[width=\textwidth]{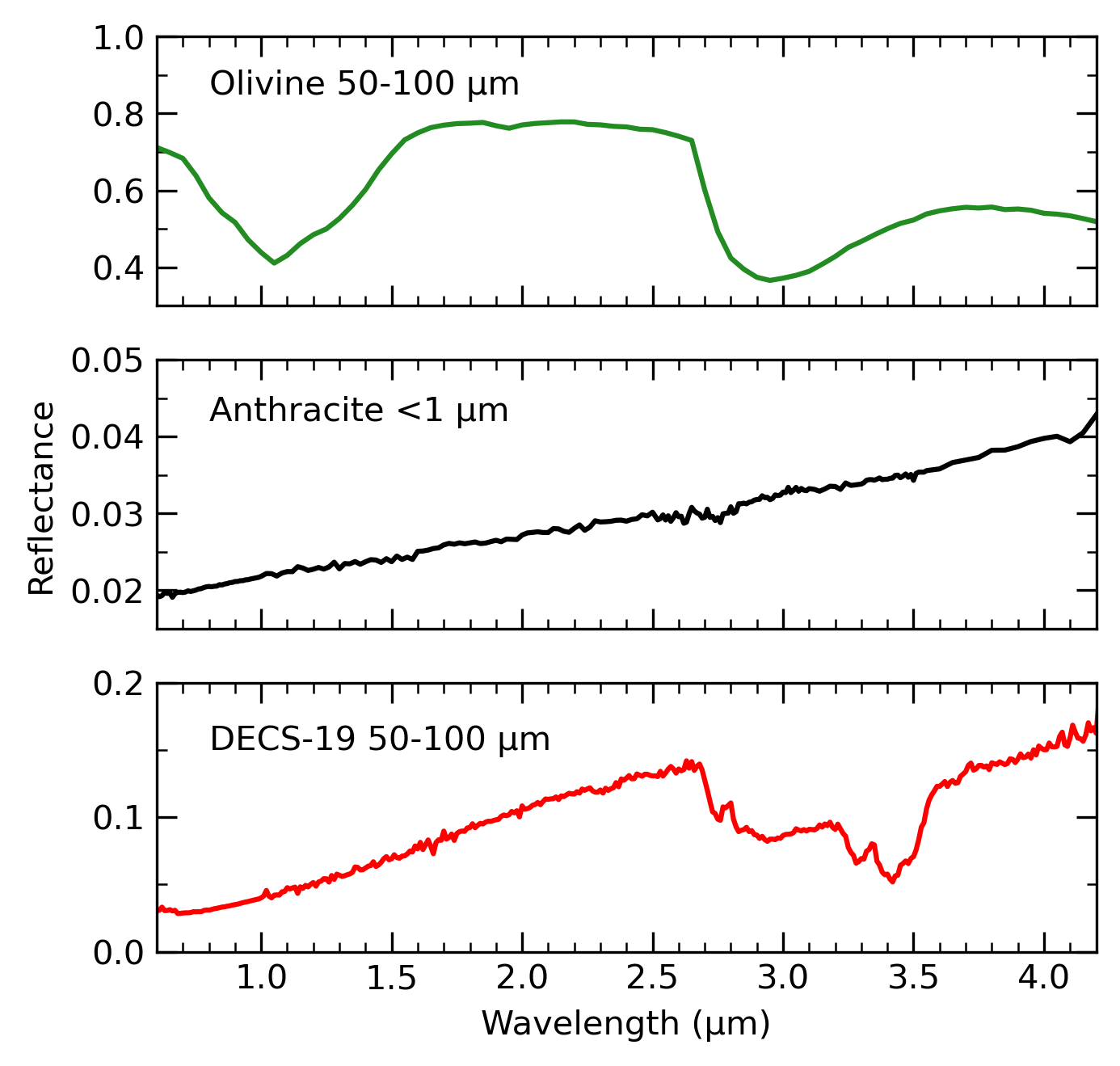}}
         \caption{}
         \label{fig:spectres_endmembers_simulant}
     \end{subfigure}
     \hfill
     \begin{subfigure}[b]{0.49\textwidth}
         \centering
         \resizebox{\hsize}{!}{\includegraphics[width=\textwidth]{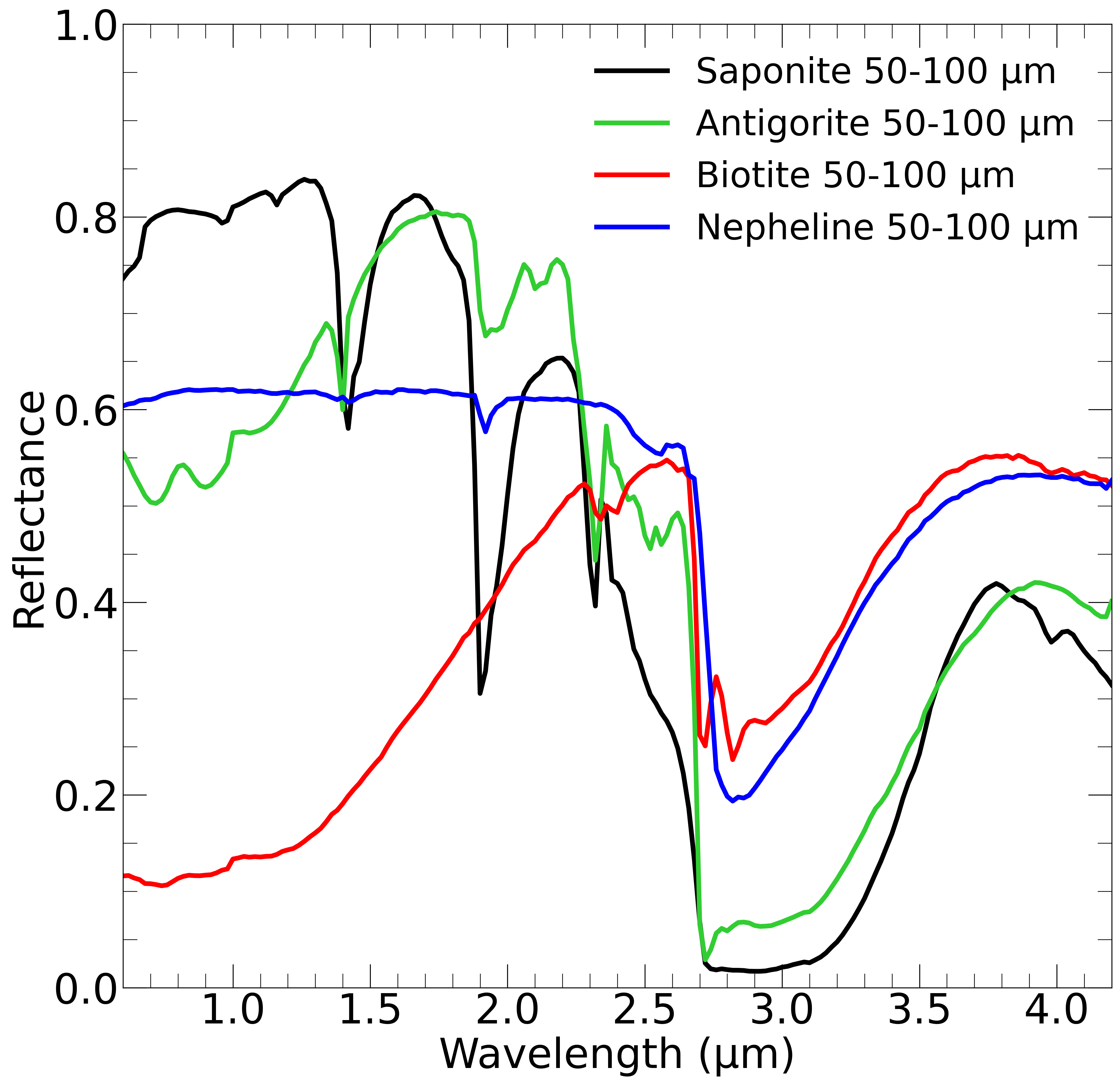}}
         \caption{}
         \label{fig:spectres_phyllosilicates_VNIR}
     \end{subfigure}
     \caption{(a) Bidirectional reflectance spectra in the visible and near-infrared of the endmembers used in the Phobos simulant proposed in \cite{Wargnier_2023_2}. The 3 µm-band visible in the olivine and DECS-19 spectra is mainly due to adsorbed atmospheric water in the sample powder. Note that the reflectance scale is different for each of the three samples. (b) Bidirectional reflectance spectra in the visible and near-infrared of some of the phyllosilicates used in this study. Phyllosilicates spectra were made on powders with the same grain size range for the different endmembers using the SHADOWS instrument.}
     \label{fig:VNIR_spectra}
\end{figure*}

The olivine spectrum (Fig. \ref{fig:spectres_endmembers_simulant}) exhibits a typical 1 µm absorption band due to iron ions transition \citep{Cloutis_2012a, Sultana_2021} and also an important feature at 3.0 µm. This feature is attributed to the adsorption of atmospheric water in olivine powder. The attribution was confirmed by the total removal of the band when measuring under vacuum (see Fig. \ref{fig:vacuum_effect}). The reflectance of olivine is quite high (70\%). Anthracite is featureless and red-sloped from 0.6 to 4.2 µm with a reflectance of 2\% at 0.6 µm and 4\% at 4.2 µm. Titan tholins are relatively bright, in particular between 1 and 3 µm, with a reflectance up to roughly 70\%, and exhibit a deep feature due to N-H stretching modes at 3 µm which has been particularly discussed and studied in \cite{Wargnier_2023_1}. DECS-19 exhibits C-H stretching modes at 3.28 and 3.42 µm (Fig. \ref{fig:spectres_endmembers_simulant}. The absorption band at 3.28 µm is assigned to aromatic groups. The 3.42 feature is attributed to C-H stretching modes of methylene (CH$_2$) aliphatic structures \citep{Moroz_1998, Vinogradoff_2021}. From the information provided by the Penn State Coal Sample Bank, DECS-19 has a H/C of 0.6 which corresponds to a composition with 85\% of carbon. Such a H/C is notably in agreement with the H/C on asteroid Bennu, which would be between 0.3 and 0.6 \citep{Kaplan_2021}. Although coals could be representative of the composition of a small body such as Phobos, they were chosen for their spectral properties and their aromatic/aliphatic features. DECS-19 presents another absorption band near 3 µm which is due to water adsorption in the coal powder (see Fig. \ref{fig:vacuum_effect} in Appendix). It is also red-sloped between 0.6 and 4.2 µm, and the reflectance varies from 2.5\% at 0.6 µm to 17\% at 4.2 µm. \newline
Pertaining to phyllosilicates (Fig. \ref{fig:spectres_phyllosilicates_VNIR}), antigorite shows overtone features between 0.6 and 2.5 µm and a reflectance on average of 0.6 in this wavelength range. A deep 2.7 µm O-H feature can be observed, accompanied by a molecular water absorption band at 3.0 µm (Fig. \ref{fig:vacuum_effect}). Biotite, from 0.6 to 2.5 µm, exhibits a red slope and is darker than antigorite with a reflectance of 0.1 at 0.6 µm. Reflectance of biotite becomes higher than antigorite after 2.5 µm and up to 4.2 µm. Biotite shows also a 2.7 µm absoption band and a 3.0 µm feature (Fig. \ref{fig:vacuum_effect}). Nepheline spectrum is flat from 0.6 to 4.2 µm with a reflectance of 0.6 at 0.6 µm. Because nepheline is a feldspathoid, the 2.7 µm feature is not present in the spectrum. But a 3 µm band can be observed, due to molecular water and/or O-H defects due to the natural origin of the nepheline. Saponite is relatively bright especially for $\lambda <$ 1.3 µm with a reflectance of 0.8. It presents also deep absorption feature at 1.45 (OH), 1.9 (OH/H$_{2}$0), and 2.3 µm (Mg-OH) \citep{Cloutis_2012a}. The 2.7 µm band is visible and is associated with a deep and large 3 µm trough. This is because saponite, a clay mineral, absorbs a significant amount of atmospheric water. Despite using a calcium-rich saponite, a faint feature was observed in the 0.65-0.7 µm regions, possibly due to Fe$^{2+}$-Fe$^{3+}$ charge transfer \citep{Cloutis_2012a, Cloutis_2012b}. The antigorite spectrum also exhibits this feature, with a larger band and a center position around 0.7-0.75, also in this case due to iron charge transfer, as noted by \cite{Cloutis_2012a}. An additional feature is also visible in the antigorite spectrum at 0.9 µm, probably due to an octahedral Fe$^{2+}$ crystal field transition \citep{Cloutis_2012a}.

\subsubsection{In the MIR}
Mid-infrared spectra of olivine, anthracite, and DECS-19 were already discussed in details in \cite{Wargnier_2023_2}. None of these materials showed spectroscopic features similar to Phobos in the MIR. Olivine shows a CF at $\sim$9 µm, anthracite at $\sim$7.9 µm, and DECS-19 at $\sim$7 µm. \par
It should be noted that the measurements were taken under standard atmospheric conditions and at room temperature. Although this may affect the VNIR spectrum, the effects on the MIR spectrum are more pronounced and therefore not entirely representative of the conditions on Phobos. This could result in variations, such as differences in feature position or intensity \citep{Salisbury_1989, Salisbury_1991, Cooper_2002}. However, it is important to note that the approximation of Kirchoff's law to obtain emission spectra may also potentially modify the band position and intensity. In this work, MIR spectra are used as indicators to fit and compare the Phobos spectrum, but variations may occur under different experimental conditions. \par
\begin{figure}
\centering
\resizebox{10cm}{!}{\includegraphics{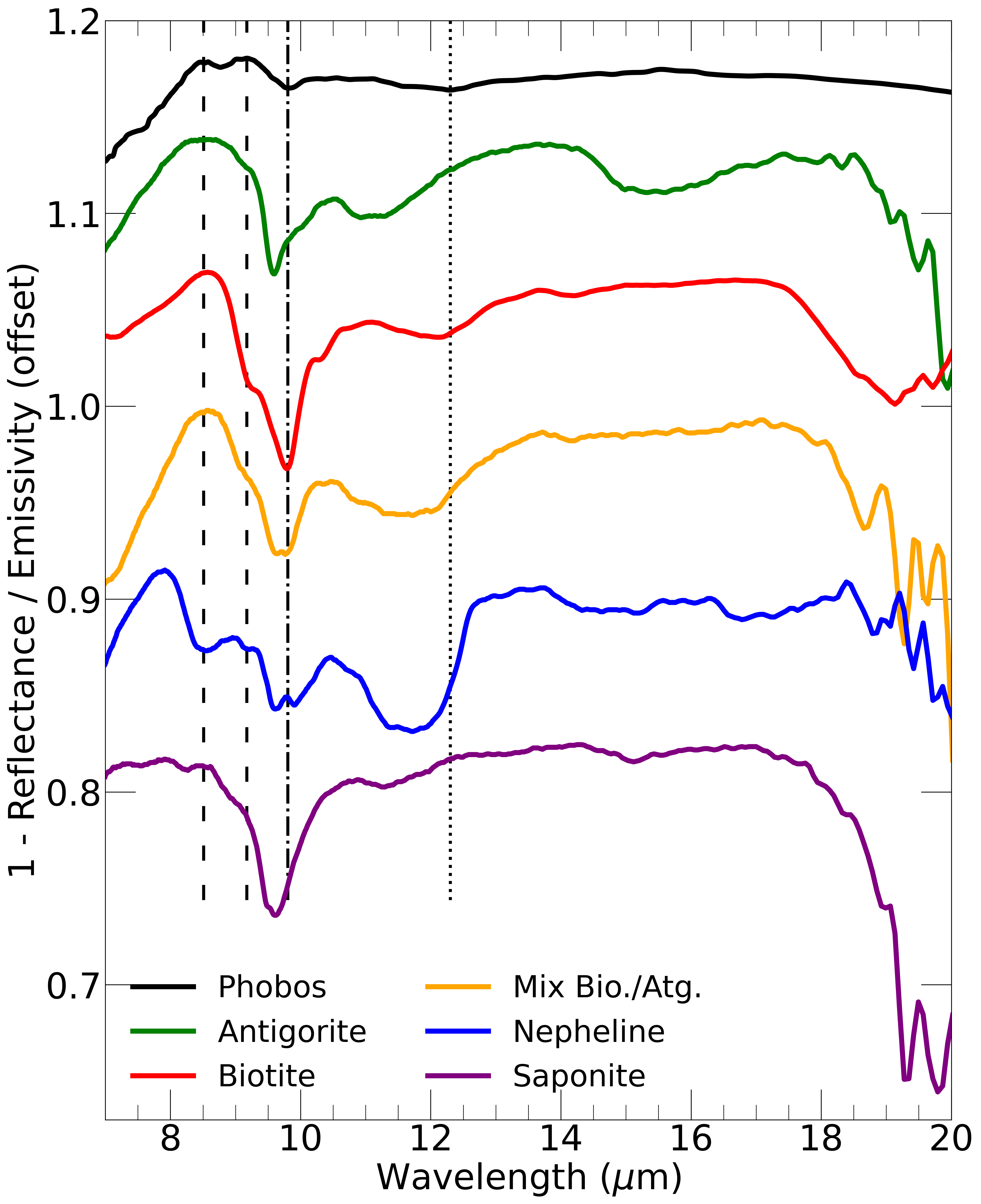}}
\caption{Bidirectional 1-reflectance spectra in the mid-infrared of phyllosilicates used in this study compared to the TES red unit emissivity spectrum of Phobos \citep{Glotch_2018}. The "Mix Bio./Atg." is a 50:50 mixture of biotite and antigorite. Grain size of the phyllosilicates is 50-100 µm (Table \ref{tab:composition_mixtures}). The vertical dashed line indicates the positions of the Christiansen features of the Phobos spectrum. The vertical dash-dotted and dotted line show respectively the positions of the Phobos restrahlen band and of the transparency feature. Note that the biotite spectrum was re-scaled and divided by a factor of 4, for clarity. Positions of the CF, RB, and TF of the biotite are in relatively good agreement with the Phobos mid-infrared features.} 
\label{fig:spectres_phyllosilicates_MIR}
\end{figure}

Several phyllosilicates spectra were investigated in the MIR (Fig. \ref{fig:spectres_phyllosilicates_MIR}) and the choice of the best phyllosilicates endmembers for the following measurements, was based on the comparison with TES red unit spectrum of Phobos \citep{Glotch_2018}. CF of nepheline is also shifted toward smaller wavelength compared to the TES spectrum, but the overall spectrum better matches the Phobos spectrum. Saponite and antigorite spectra have a CF respectively that match well the first CF of the Phobos spectrum and their RB is close to the position of the band observed in Phobos spectrum. The best match in the MIR for a single endmember is the biotite. Positions of the three important features (CF, RB, and TF) match the position of the same features in the Phobos spectrum. As suggested by \cite{Giuranna_2011}, a linear mixture of biotite and antigorite in equal quantities was also measured. However, looking only at the positions of CF, RB, and TF, pure biotite is more in agreement with the Phobos observation. Note that determining the mineralogical composition with a double CF can be complicated. This type of complex CF may result from the combination of two distinct endmembers that can be spectrally resolved, and therefore manifest themselves as two distinct peaks \citep{Salisbury_1991}.

\subsection{Detection of aliphatic/aromatic bands in Phobos spectral simulants} \label{sect:det_orga}

\begin{table}
	\centering
	\caption{Estimation of hydrocarbon CH$_x$ groups in each mixtures for organics detectability study.}
	\label{tab:detectability_orga}
	\resizebox{10cm}{!}{
	\begin{tabular}{cccccc} 
		\hline
		\textbf{Mixtures} & \textbf{DECS-19} & \textbf{DECS-19} & \textbf{Carbon} & \textbf{Aliphatic} & \textbf{Aromatic} \\
        & (vol.\%) & (wt.\%) & (wt.\%) & (wt.\%) & (wt.\%) \\
		\hline
        CH-rich-5 & 5 & 3.08 & 2.64 & 0.42 & 2.21 \\
        CH-rich-10 & 10 & 6.28 & 5.37 & 0.87 & 4.51 \\
        CH-rich-15 & 15 & 9.62 & 8.23 & 1.33 & 6.91 \\
        CH-rich-20 & 20 & 13.11 & 11.22 & 1.81 & 9.41 \\
        CH-rich-25 & 30 & 20.55 & 17.59 & 2.83 & 14.75 \\
        \hline
	\end{tabular}
	}
\end{table}

\begin{figure*}
     \centering
     \begin{subfigure}[b]{0.49\textwidth}
         \centering
         \resizebox{\hsize}{!}{\includegraphics[width=\textwidth]{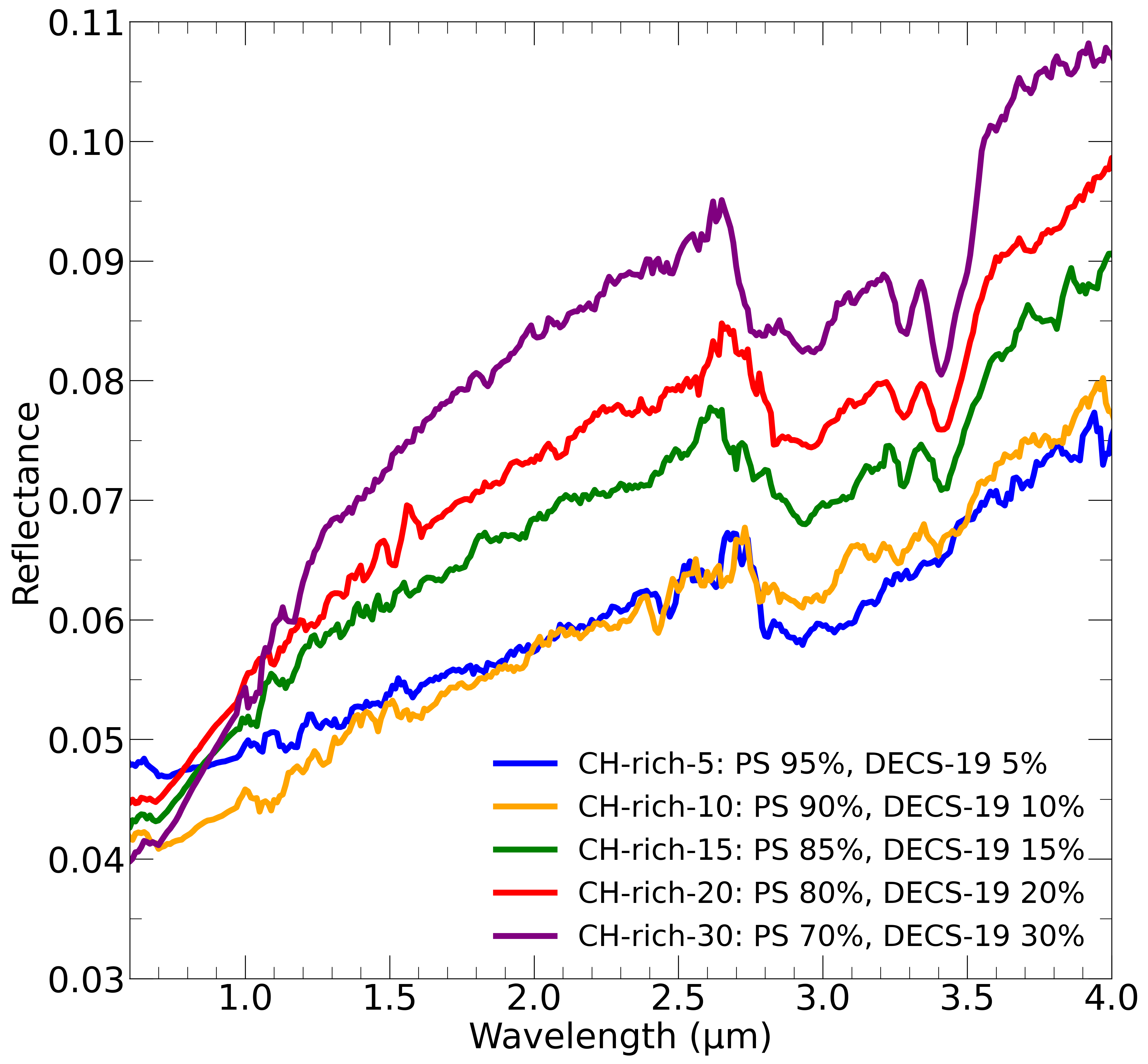}}
         \caption{}
         \label{fig:SIM1_orga_spec}
     \end{subfigure}
     \hfill
     \begin{subfigure}[b]{0.49\textwidth}
         \centering
         \resizebox{\hsize}{!}{\includegraphics[width=\textwidth]{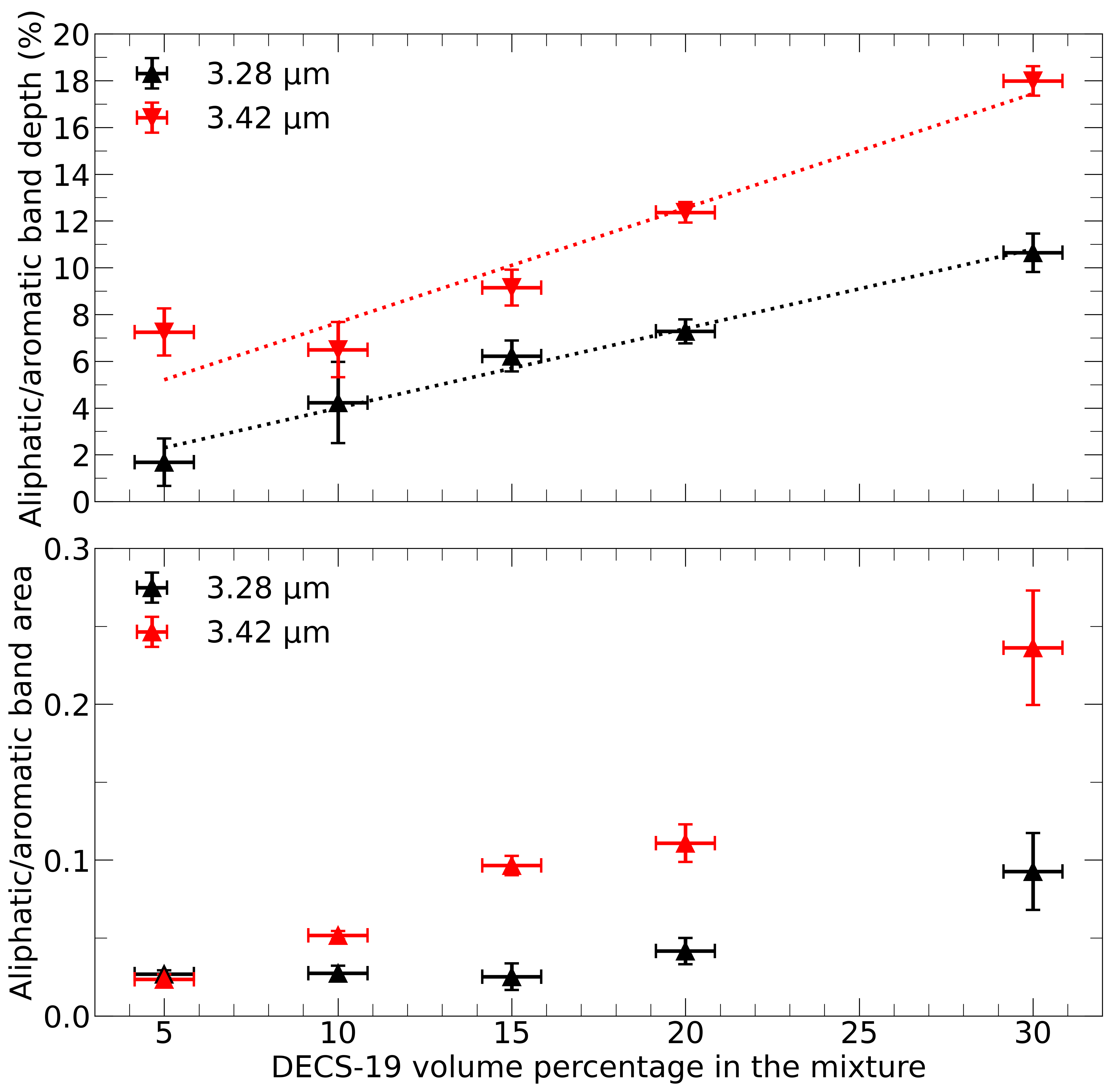}}
         \caption{}
         \label{fig:SIM1_orga_det}
     \end{subfigure}
     \caption{(a) Reflectance spectra of the mixtures CH-rich-5, -10, -15, -20, and -30, based on the PSD simulant with addition of various quantities of organics, from 5 to 30 vol.\%. Spectra are smoothed on these figures using a moving average of 3 points. (b) Top: Band depth of the 3.28 and 3.42 µm features as a function of the quantity of the organic compounds in PSD. Bottom: Band area of the 3.28 and 3.42 µm features as a function of the quantity of the organic compounds in PSD. Band areas are computed respectively from 3.2 µm to 3.35 µm and from 3.35 µm to 3.6 µm}
     \label{fig:spectre_SIM12_orga}
\end{figure*}

The MIRS spectrometer onboard MMX will observe for the first time the 3.4 µm region of the Phobos spectrum with sufficient SNR. Using ground-based observations, \cite{Takir_2022} have shown the absence of C-H stretching mode absorption bands but only for band depth larger than 15-20\% due to the uncertainties of the measurement. However, organics on Phobos could be altered by space weathering and hence, less detectable. It also important to note that other small bodies have shown features in this wavelength range related to organic materials, such as Bennu \citep{Kaplan_2021}, Ryugu \citep{Yada_2022, Pilorget_2022, Hatakeda_2023}, 67P \citep{Raponi_2020}, Ceres \citep{DeSanctis_2017, DeSanctis_2019} and in insoluble organic matter (IOM) of meteorites \citep{OrthousDaunay_2013, Kaplan_2019}. Presence or organics in primitive asteroids are particularly discussed in \cite{Hromakina_2022}. They found no 3.4 µm signature in the spectrum of the D-type Bononia \citep{Takir_2012}, despite the good signal-to-noise ratio, but Jupiter trojans asteroids have shown signatures of organics \citep{Brown_2016, Wong_2023}. In this context, it appears important to study beforehand the detection of this type of organic feature to interpret future MIRS observations. We added organic compound (DECS-19) in different proportions from 5 to 30 volume percent in PSD. Based on the vitrinite reflectance (1.71\%) of DECS19 and the relationship with aliphatic groups obtained through linear regression of the data from \cite{Wei_2018}, we were able to derive the contribution of aliphatic and aromatic carbon. Our analysis revealed that 16\% of carbon is involved in aliphatic structures and 84\% in aromatic structures. This is consistent with the results of the RMN measurements on DECS-19 presented in \cite{Vinogradoff_2021}. Therefore, it is possible to estimate the minimum quantities of aliphatic and aromatic carbon structures required for detection by IR spectroscopy (Table \ref{tab:detectability_orga}). For instance, the addition of 10 vol.\% of DECS-19 to the PSD mixture results in the incorporation of 5.37 wt.\% of carbon, with 4.51 wt.\% corresponding to aromatic carbon structures and 0.87 wt.\% to aliphatic structures. The two organic features at 3.28 µm and 3.42 µm are clearly observed (Fig. \ref{fig:SIM1_orga_spec}). We study, here, the evolution of the absorption band depth as the proportion of DECS-19 increases, to determine the detectability of 3.4 µm organics.\newline
Results of this investigation are shown in Fig. \ref{fig:SIM1_orga_det}. The depth of the 3.28 µm and 3.42 µm features increases with the DECS-19 volume percentage quantity. Using a linear fit, we find a slope of 0.34 $\pm$ 0.03 \%/µm with a zero-point of 0.60 $\pm$ 0.58 \% for the 3.28 µm feature. For the 3.42 µm absorption band, we compute a slope of 0.49 $\pm$ 0.07 \%/µm and 2.76 $\pm$ 1.37 \% for the zero-point. \par
These results can be compared with those presented in \cite{Wargnier_2023_1}, obtained using tholins as an organics source. The 3 µm N-H feature linked to the presence of organics was much bigger than the aliphatic/aromatic bands in this work and the limit of detectability was reached for 5 vol.\% of tholins in the simulant. For the same geometry, the N-H band depth is around 25\%. But the bands studied in this work, are more representative of what we expect on the Phobos surface in the case of the captured asteroid hypothesis.

\begin{figure}
\centering
\resizebox{8cm}{!}{\includegraphics{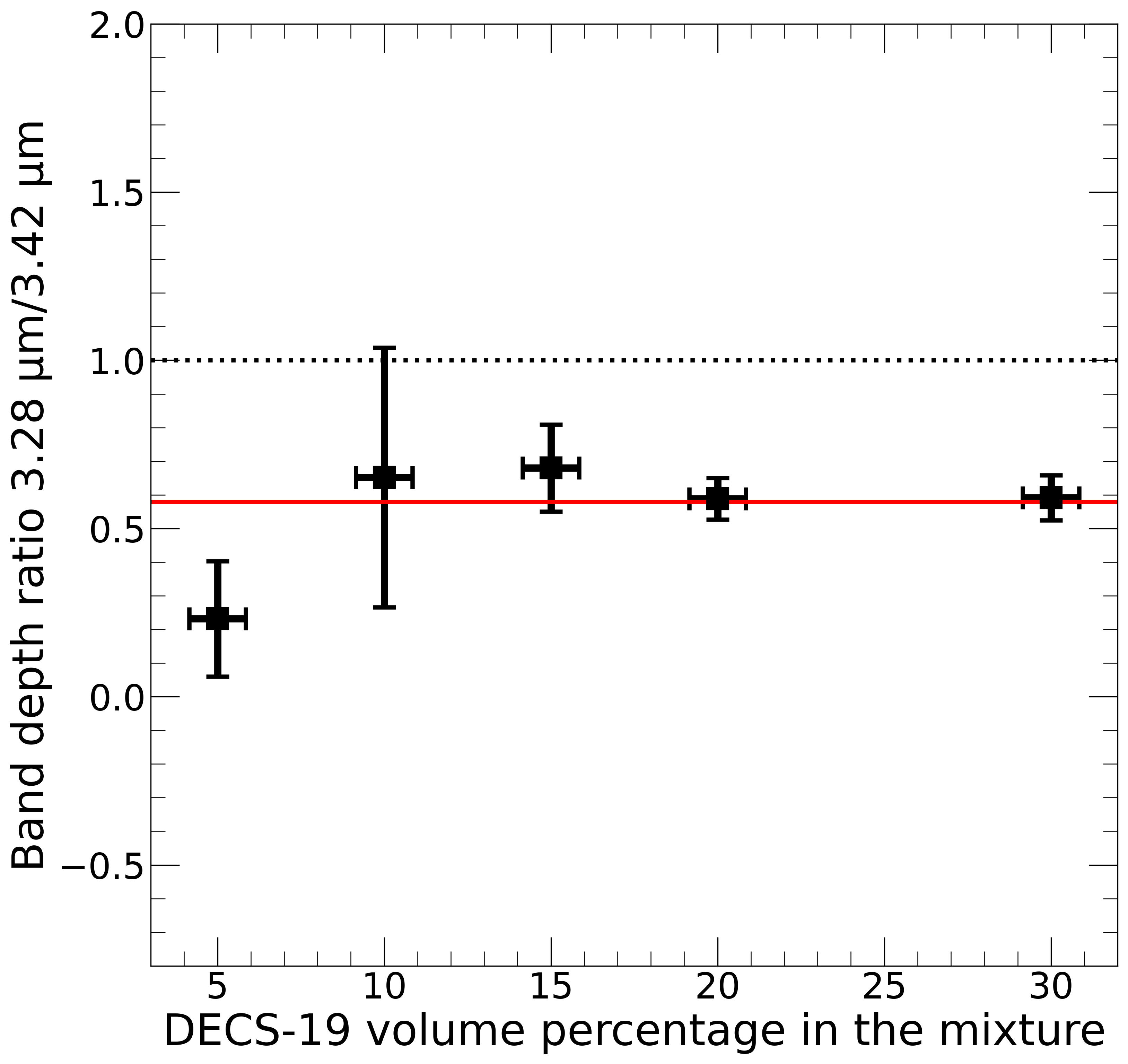}}
\caption{Band depth ratio of the 3.28 µm feature with the 3.42 µm feature. The red horizontal line represents the best fit (0.579 $\pm$ 0.002), taking into account the uncertainties of the data points. Dashed horizontal line represents the value where the band depth of the 3.28 µm absorption band could be equal to the band depth of the 3.42 µm feature.}
\label{fig:ratio_band_depth}
\end{figure}

Finally, Fig. \ref{fig:SIM1_orga_det} highlights the quite high values of band depth for the 3.42 µm feature for organic-rich composition. The maximum value obtained -- for the 3.42 µm absorption band and for 30\% of DECS-19 in PSD -- is a bit larger than 18\%. However, the band depth is strongly dependent on the organic quantity in the simulant. \\
Fig. \ref{fig:ratio_band_depth} represents the ratio of the 3.28 µm absorption band depth to the 3.42 µm band depth. We observed no significant trend in this ratio when the quantity of organics increased. It seems that the ratio is quite constant within the error bars. Values of the ratio are nearly 0.55 for both simulants. The band at 3.42 µm is therefore about twice as deep as the one at 3.28 µm. However, as discussed previously (Sect. \ref{sect:bd_compute}), the 3.28 µm is a bit underestimated compared to the 3.42 µm due to the presence of the 3.0 µm molecular water feature. 

\subsection{Detectability of hydrated minerals} \label{sect:det_hydmin}

\begin{table}
	\centering
	\caption{Estimation of hydroxyl OH groups in each mixtures for hydrated minerals detectability study.}
	\label{tab:detectability_det_hydmin}
	\begin{tabular}{cccc} 
		\hline
		\textbf{Mixtures} & \textbf{Antigorite} & \textbf{Antigorite} & \textbf{OH} \\
        & (vol.\%) & (wt.\%) & (wt.\%) \\
		\hline
        hydmin-rich-1 & 1 & 0.9 & 0.24 \\
        hydmin-rich-3 & 3 & 2.7 & 0.73 \\
        hydmin-rich-5 & 5 & 4.5 & 1.22 \\
        hydmin-rich-10 & 10 & 9.1 & 2.46 \\
        hydmin-rich-15 & 15 & 13.7 & 3.70 \\
        hydmin-rich-20 & 20 & 18.3 & 4.94 \\
        \hline
	\end{tabular}
\end{table}

\begin{figure*}
     \centering
     \begin{subfigure}[b]{0.49\textwidth}
         \centering
         \resizebox{\hsize}{!}{\includegraphics[width=\textwidth]{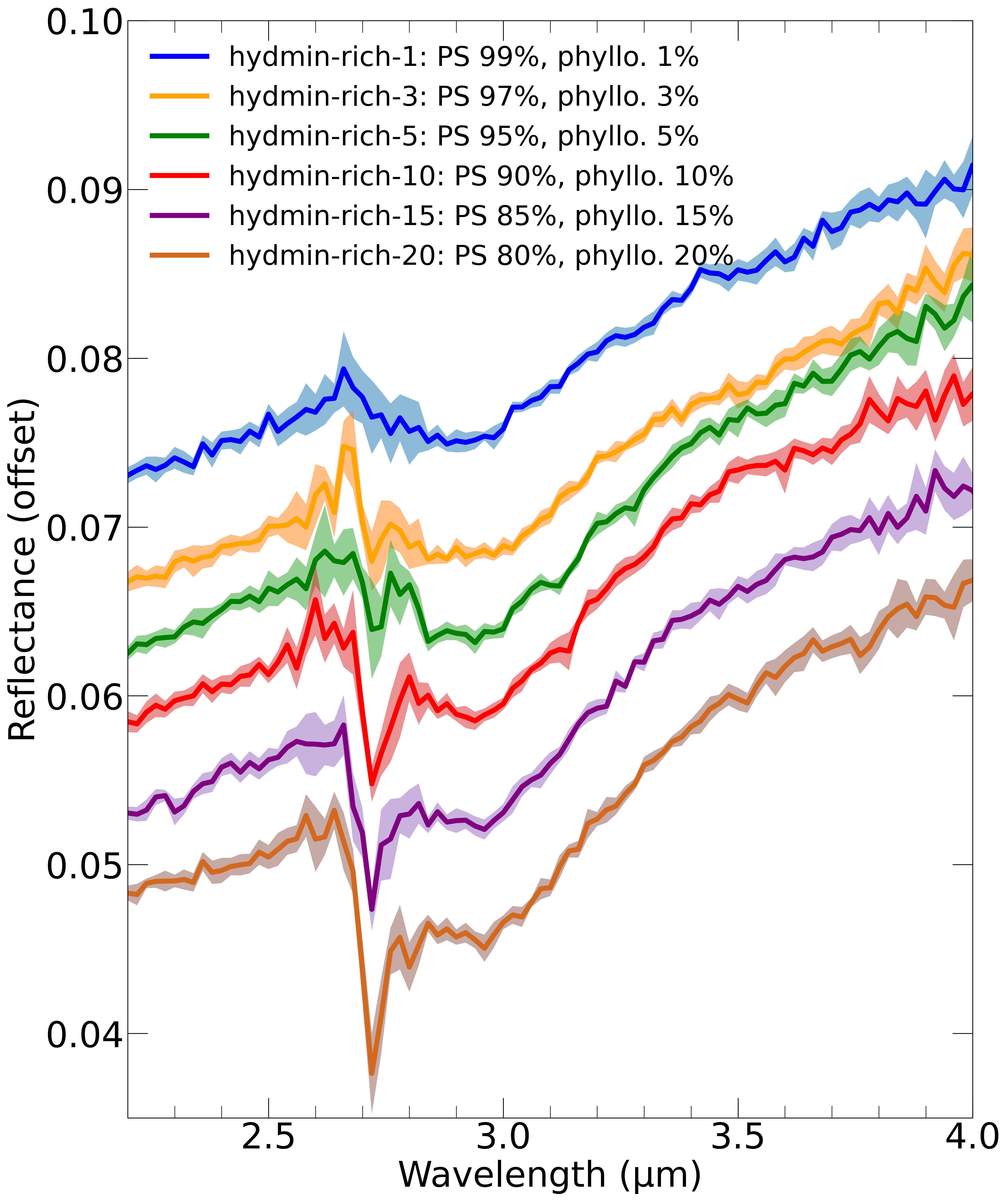}}
         \caption{}
         \label{fig:SIM1_hydramin_spec}
     \end{subfigure}
     \hfill
     \begin{subfigure}[b]{0.49\textwidth}
         \centering
         \resizebox{\hsize}{!}{\includegraphics[width=\textwidth]{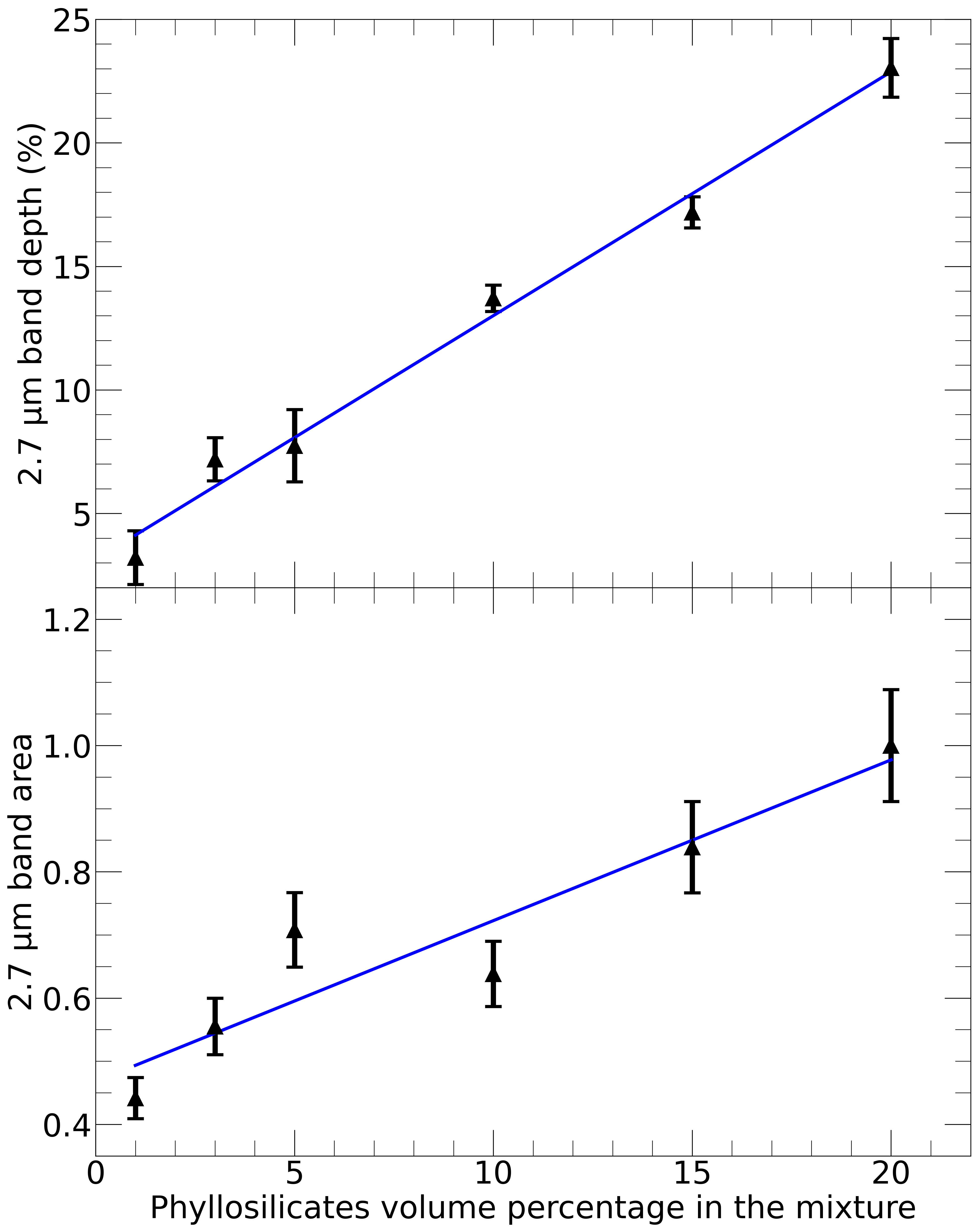}}
         \caption{}
         \label{fig:SIM1_hydramin_det}
     \end{subfigure}
     \caption{(a) SHADOWS reflectance spectra of the mixtures hydmin-rich-1, -3, -5, -10, -15, and -20, based on the PSD simulant with addition of various quantities of phyllosilicates, from 1 to 20 vol.\%. The shaded areas correspond to the uncertainties of the measurements. (b) Top: Band depth of the 2.7 µm feature as a function of the quantity of the hydrated minerals in PSD. Bottom: Band area of the 2.7 µm feature as a function of the quantity of the hydrated minerals in PSD. Band areas are computed from 2.6 µm to 3.4 µm, hence by taking into account the 3 µm feature. Band areas are normalized to the maximum value obtained for 20 vol.\% in the simulant.}
     \label{fig:spectre_SIM12_hydmin}
\end{figure*}

Because hydrated minerals can be expected for Phobos but were never unambiguously detected, our aim is also to investigate the detectability of hydrated minerals in a Phobos simulant, in preparation for future MIRS observations. We decided to use the antigorite as the phyllosilicates part for the study of detectability of the 2.7 µm feature characteristic of hydrated minerals. As for organics (Sect. \ref{sect:det_orga}), using the Phobos simulant PSD from \cite{Wargnier_2023_1}, we added the phyllosilicate in various quantities from 1 to 20 vol.\% (Fig. \ref{fig:SIM1_hydramin_spec}). The composition and density of the material were also considered in order to express the phyllosilicate quantities in terms of OH groups, which is an important parameter for the detectability of the 2.7 µm band. For example, 10 vol.\% of antigorite in the mixture corresponds to 2.46 wt.\% of hydroxyl groups (Table \ref{tab:detectability_det_hydmin}). This approach allows for the straightforward extension of this work to other phyllosilicates with varying OH compositions. We used the SHADOWS instrument for this study, as the spectral resolution is close to the MIRS resolution (i.e., 20 nm). We found, as expected, that bands depth increase with increasing quantity of phyllosilicates (Fig. \ref{fig:SIM1_hydramin_det}). A linear fit of band depth as a function of phyllosilicates quantity gives a zero-point of 3.13 $\pm$ 0.60 \% and a slope of 0.99 $\pm$ 0.05 \%/µm. Hence, the 2.7 µm antigorite band increases particularly rapidly as the quantity of hydrated minerals increases. The O-H stretching mode feature appears to be narrow and intense. The minimum of the band is always centered at 2.72 µm. The band is generally associated with a 3 µm feature due to water adsorption and/or O-H defect in the natural sample of minerals. As the 2.7 µm feature is narrow, the limited resolution of the instrument can lead to a non-complete band. This effect is visible for the hydmin-rich-5 mixture (Fig. \ref{fig:SIM1_hydramin_spec}). However, the band is still clearly visible and do not have effect on the detectability. This effect will be important to take into account for quantification purpose. Band depth ranges from 3\% to 23\%. For a same volume fraction, hydrated mineral feature are two times deeper than the organics features at 3.28 and 3.42 µm.

\subsection{Update of the Phobos simulant}
We tried to reproduce the main properties of the Phobos spectrum using spectra from the visible to the mid-infrared (see also \citealt{Wargnier_2023_2}), using the spectral properties observed at different geometries of observation. Table \ref{tab:composition_mixtures} describes the mixtures made for this study. We compared the different mixtures in terms of VNIR spectral slope and reflectance and MIR absorption bands positions. Spectral parameters of a laboratory simulant in the VNIR were already studied previously \citep{Wargnier_2023_2}. The best-fit mixture compared to CRISM spectrum of Phobos \citep{Fraeman_2012} was found to be a mixture of olivine (60 vol.\%, 50-100 µm), anthracite (20 vol.\%, $<$ 1 µm) and DECS-19 (20 vol.\%, $<$ 50-100 µm). We chose to use this VNIR Phobos simulant as the basis of the new investigations. As discussed previously (Sect. \ref{sect:intro} and \ref{sect:det_hydmin}), phyllosilicates have been found to be a possible explanation of the features observed in the MIR. Hence, phyllosilicates were added to a mixture of olivine, anthracite, and DECS-19. According to pure endmembers spectra (Sect. \ref{sect:results_spec}), saponite, biotite, antigorite, and nepheline were selected to match the Phobos MIR spectrum. We found that a mixture composed of biotite and antigorite is indeed a good simulant for the MIR spectrum of Phobos as shown in \cite{Giuranna_2011}. Positions of CF, RB, and TB are consistent with the Phobos spectrum. \par
\begin{figure}
\centering
\resizebox{10cm}{!}{\includegraphics{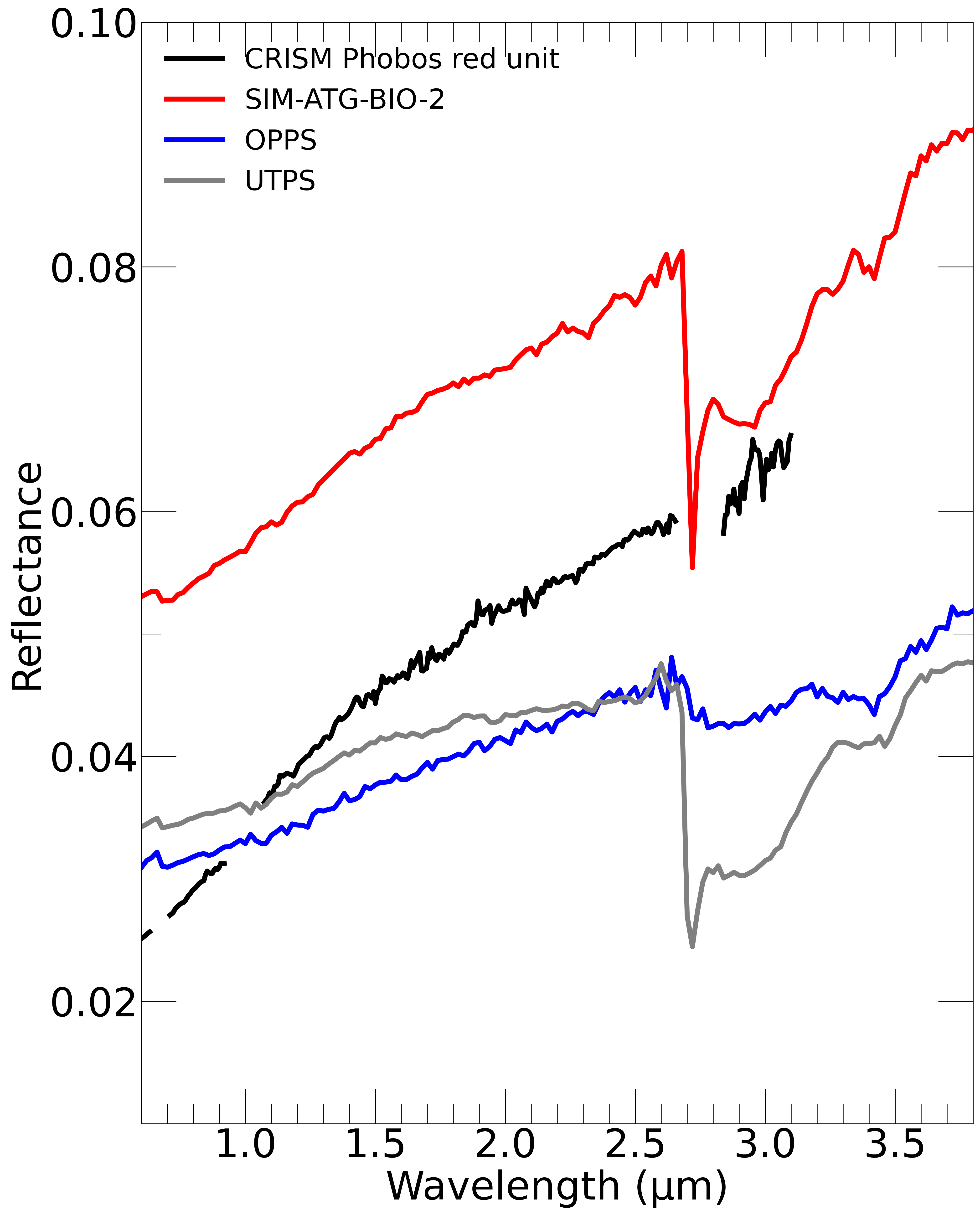}}
\caption{Visible and near-infrared spectra of mixtures SIM-ATG-BIO-2 and SIM-SAP-1 (OPPS) compared to the Phobos red unit spectrum \protect \citep{Fraeman_2014} reprojected to laboratory geometry (i = 0$\degree$, e = 30$\degree$). Spectrum of the UTPS simulant is also show for comparison \protect \citep{Miyamoto_2021}.}
\label{fig:VNIR_simulant}
\end{figure}

However, the VNIR spectrum of this simple mixture is not in agreement with the Phobos spectrum in this wavelength range. Mixtures that match well Phobos spectra from the VNIR to the MIR are quite complicated mixtures composed of 4 to 5 endmembers (SIM-ATG-BIO, SIM-NEP, SIM-ATG-SAP, and SIM-SAP). Looking at VNIR spectra of the different mixtures (Fig. \ref{fig:VNIR_simulant} and Table \ref{tab:mix_params}), we observed that most of them are quite good in terms of spectral slope except the SIM-ATG-BIO-1 mixture. Spectral slopes of SIM-ATG-BIO-1 are too low compared to Phobos spectral slope. SIM-ATG-BIO-1 does not contain DECS-19. As seen in \cite{Wargnier_2023_2}, DECS-19 is an important endmembers for the spectral slope in the VNIR and few percent can change drastically the slope. The steeper slopes in the visible appear to be linked with the presence of nepheline and a high quantity of DECS-19 (20 vol.\%). From 1.5 to 2.4 µm, the slopes show less variations and are comprised between 1.13 \%/100nm and 2.82 \%/100nm. These last values are close to the slope value in the same wavelength range found in the Phobos CRISM spectrum \citep{Fraeman_2012}. \newline
Reflectance of the different mixtures is comprised between 0.03 and 0.06 at 600 nm whereas the reflectance of the Phobos spectrum at this wavelength is about 0.025. Therefore, for most of the mixtures, the reflectance is a bit too high compared to Phobos. Using these endmembers it was difficult to decrease the reflectance level without any significant changes for the spectral slope. All mixtures except SIM-ATG-BIO-1 exhibit two organics bands at 3.28 and 3.42 µm due to DECS-19. These bands are more or less important depending on the quantities of DECS-19 and of anthracite. Mixtures containing phyllosilicates present the 2.7 µm O-H feature. Mixtures with nepheline instead of phyllosilicates do not have this absorption band but show a 3 µm feature. Except for O-H and C-H features, other bands that can be seen in pure endmembers spectra (Fig. \ref{fig:spectres_phyllosilicates_VNIR}) are completely removed.

\begin{figure}
\centering
\resizebox{10cm}{!}{\includegraphics{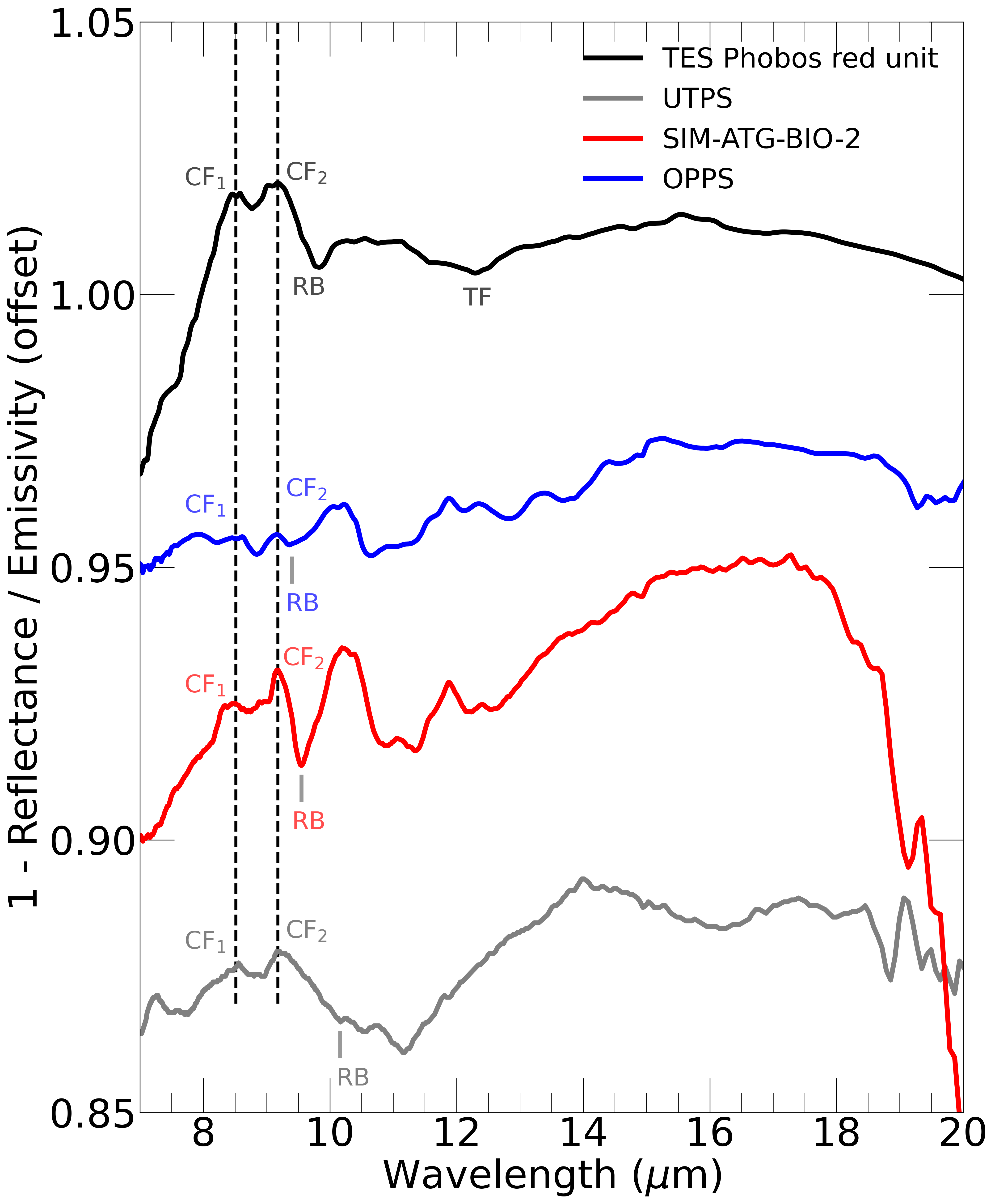}}
\caption{Mid-infrared spectra of mixtures SIM-ATG-BIO-2 and SIM-SAP-1 (OPPS) compared to TES spectrum of Phobos from \protect \cite{Glotch_2018}. Spectrum of the UTPS simulant \protect \citep{Miyamoto_2021} is also show for comparison. The UTPS spectrum was measured on a powder composed of grains $<$125 µm. The two vertical dashed line represent the position of the Christiansen features of the TES Phobos spectrum at 8.59 µm and 9.17 µm. Spectra are offset for clarity.}
\label{fig:MIR_simulant}
\end{figure}

In the mid-infrared range, mixtures reveal different behaviors depending on the composition (Fig. \ref{fig:MIR_simulant}), in particular looking at the CF, RB, and TF. Positions of the three main features for the mixtures are given in Table \ref{tab:mix_params}. Mixtures containing nepheline are flatter than the others with antigorite, biotite, and saponite; and CF are not well defined but are in the range of the double CF observed in the TES Phobos spectrum \citep{Glotch_2018}. Mixtures containing saponite and the mixture of antigorite and biotite present an interesting double CF at the position of the CF given by the TES Phobos spectrum. Restrahlen bands positions do not change within uncertainties with the composition of the mixture. Indeed, this feature is mainly related to the silicate class and therefore all phyllosilicates show a close RB in position and shape. The shape and contrast of the RB could also be modified by grain size \citep{Salisbury_1991}. However, the grain sizes used in the mixtures are the same for a given endmember. RBs ($\sim$ 9.5 µm) are systematically shifted toward smaller wavelengths compared to the Phobos spectrum ($\sim$ 9.8 µm). Interestingly, using ion irradiation to simulate space weathering, \cite{Brunetto_2020} noticed a shift of the RB toward larger wavelengths for the irradiated surface that could explain the difference between fresh and altered samples. For the different mixtures, TFs were found in the range 12.09 - 12.80 µm. \newline
Although a quantitative study cannot be made of the emissivity value, as the spectra were transformed from reflectance to emission using Kirchoff's law, it is interesting to note that the various mixtures present a high emissivity of around 0.95 at 8 µm. This order of magnitude of emissivity is in agreement with the emissivity at the same wavelength measured on Phobos by TES of approximately 0.99. \par
According to the spectral slopes, the reflectance, the different positions of the features, and of the overall shape of the spectrum, SIM-ATG-BIO-2 and SIM-SAP-1 were chosen as the best Phobos spectral simulants from the VNIR to the MIR. They appear to be a great compromise between the different spectral parameters described above. It should be noted that we have used here the MIR spectrum given in \cite{Glotch_2018}, but that the Phobos MIR spectra of \cite{Giuranna_2011} show significant variations with the latter. It is therefore difficult to draw definitive conclusions in this spectral range. Although SIM-ATG-BIO-2 has spectral properties more similar to those of Phobos, SIM-SAP-1 was chosen due to its composition, which includes saponite. This mixture is interesting because endmembers are likely more representative, particularly for martian-like materials on the Phobos surface (giant impact hypothesis). It enables exploration of a composition that is very different from the UTPS-TB simulant. \par
The new Phobos simulant SIM-SAP-1 will be called OPPS (Observatory of Paris Phobos Simulant) in the following for the sake of clarity. The mineralogical composition of the OPPS is presented in Table \ref{tab:composition_mixtures}, the bulk chemical composition from EDX analysis in Table \ref{table:bulk_chemical_compo}, and the SEM images in Fig. \ref{fig:SEM_images}. \newline

\begin{table*}
	\centering
	\caption{Spectral parameters of the spectroscopic simulant mixtures in the visible and near-infrared and positions of the mid-infrared features CF, RB, and TF. Comparison with CRISM and UTPS-TB spectral parameters. CRISM and UTPS parameters are derived from spectra respectively in \protect\cite{Fraeman_2012} reprojected at the same laboratory geometry and \protect\cite{Miyamoto_2021}. A hyphen in the Table signifies that there is no data at this wavelength for the observation. Uncertainties on the spectral slope and reflectance measurements for the mixtures are due to SHADOWS uncertainties. The spectral slope is given in \%/100nm. SIM-SAP-1 is chosen in this work as the best simulant in the VNIR for Phobos (OPPS). Values for MIR features position are given with uncertainties of 0.05 µm. In this table, because Phobos has a unique RB, only the position of the first RB of the simulants is given.}
	\label{tab:mix_params}
    \resizebox{\textwidth}{!}{
	\begin{tabular}{c|ccc|ccc|ccc} 
		\hline
		\textbf{Mixtures} & & \textbf{Spectral slope} & & \multicolumn{3}{c}{\textbf{Reflectance}} & & \textbf{MIR features} & \\
        & & (in \%/100nm) & & & & & & \textbf{position (µm)} & \\
        & 720-900 nm & 1.5-2.4 µm & 3.7-4.0 µm & 600 nm & 1.8 µm & 4.0 µm & CF(s) & RB & TF \\
		\hline
		SIM-ATG-BIO-1 & 0.09 $\pm$ 0.01 & 1.29 $\pm$ 0.03 & 1.05 $\pm$ 0.04 & 0.052 $\pm$ 0.001 & 0.056 $\pm$ 0.001 & 0.074 $\pm$ 0.002 & 8.34/9.17 & 9.55 & 12.20\\
	    SIM-ATG-BIO-2 & 3.16 $\pm$ 0.01 & 1.84 $\pm$ 0.04 & 1.26 $\pm$ 0.04 & 0.053 $\pm$ 0.001 & 0.070 $\pm$ 0.001 & 0.093 $\pm$ 0.002 & 8.34/9.17 & 9.55 & 12.15\\
	    SIM-ATG-BIO-3 & 3.15 $\pm$ 0.02 & 2.42 $\pm$ 0.05 & 2.55 $\pm$ 0.07 & 0.040 $\pm$ 0.001 & 0.056 $\pm$ 0.001 & 0.082 $\pm$ 0.002 & 8.34/9.16 & 9.55 & 12.14\\
		SIM-NEP-1 & 3.49 $\pm$ 0.01 & 2.48 $\pm$ 0.05 & 1.45 $\pm$ 0.05 & 0.044 $\pm$ 0.001 & 0.062 $\pm$ 0.001 & 0.091 $\pm$ 0.002 & 9.13 & 9.61 & 12.20\\
		SIM-NEP-2 & 5.59 $\pm$ 0.02 & 2.82 $\pm$ 0.05 & 1.59 $\pm$ 0.05 & 0.038 $\pm$ 0.001 & 0.060 $\pm$ 0.001 & 0.090 $\pm$ 0.002 & 9.16 & 9.57 & 12.20\\
		SIM-ATG-BIO-4 & 2.53 $\pm$ 0.01 & 2.08 $\pm$ 0.06 & 1.04 $\pm$ 0.04 & 0.039 $\pm$ 0.001 & 0.051 $\pm$ 0.001 & 0.070 $\pm$ 0.002 & 8.34/9.18 & 9.55 & 12.09\\
		SIM-NEP-4 & 5.38 $\pm$ 0.02 & 2.76 $\pm$ 0.06 & 1.90 $\pm$ 0.06 & 0.030 $\pm$ 0.001 & 0.047 $\pm$ 0.001 & 0.071 $\pm$ 0.002 & 9.34 & 9.57 & 12.20\\
		SIM-ATG-SAP-1 & 2.05 $\pm$ 0.01 & 1.18 $\pm$ 0.02 & 1.69 $\pm$ 0.05 & 0.059 $\pm$ 0.001 & 0.073 $\pm$ 0.001 & 0.093 $\pm$ 0.002 & 8.41/9.20 & 9.54 & 12.80\\ 
  		SIM-ATG-SAP-2 & 3.42 $\pm$ 0.01 & 2.17 $\pm$ 0.03 & 1.80 $\pm$ 0.06 & 0.042 $\pm$ 0.001 & 0.059 $\pm$ 0.001 & 0.082 $\pm$ 0.002 & 8.41/9.20 & 9.54 & 12.80\\ 
        SIM-SAP-1 (OPPS) & 2.55 $\pm$ 0.01 & 2.42 $\pm$ 0.06 & -- & 0.034 $\pm$ 0.001 & 0.045 $\pm$ 0.001 & -- & 8.41/9.20 & 9.36 & --\\ 
        \hline
        UTPS-TB  & 1.53 $\pm$ 0.02 & 1.09 $\pm$ 0.04 & -0.71 $\pm$ 0.06 & 0.030 $\pm$ 0.001 & 0.037 $\pm$ 0.001 & 0.040 $\pm$ 0.001 & 8.56/9.18 & 10.16 & 11.87\\
        \hline
		\textbf{CRISM} & \textbf{8.78 $\pm$ 0.05} & \textbf{2.99 $\pm$ 0.06} & \textbf{--} & \textbf{0.025 $\pm$ 0.005} & \textbf{0.049 $\pm$ 0.003} & \textbf{--} & \textbf{--} & \textbf{--} & \textbf{--} \\
        \textbf{TES} & \textbf{--} & \textbf{--} & \textbf{--} & \textbf{--} & \textbf{--} & \textbf{--} & \textbf{8.59/9.17} & \textbf{9.82} & \textbf{12.30}\\
        \hline
	\end{tabular}
    }
\end{table*}

\subsection{The effects of the geometry of observation}
The geometry of observation is of crucial importance for remote-sensing investigations of small bodies. Several studies have shown the variation of the reflectance, slope, etc with the illumination angles, using remote-sensing data (e.g., \citealt{Clark_2002}), and from laboratory investigations (e.g., \citealt{Pommerol_2013,Yoldi_2015,Jost_2016,Jost_2017a,Jost_2017b, Potin_2022}). \\
In this work, we studied the variations of the organic and hydrated minerals band depth with phase angle. We also have a look at the evolution of the spectral slope with the geometry (i.e., phase reddening), and we mainly focused our measurements and analysis on the evolution of the reflectance with the phase (i.e., phase curve).

\subsubsection{Band depth}
\paragraph{Evolution of the organic band's depth} \mbox{} \\
Previous papers have shown that organic features band depth could be slightly modified (e.g., \citealt{Fornasier_2020, Wargnier_2023_1}) for high phase angle. Hence, to better understand parameters that can influence organic matter detection in IR spectroscopy, it is crucial to explore the effects of observation geometry, in particular on the band depth.\par
\begin{figure}
     \centering
     \resizebox{8cm}{!}{\includegraphics[width=\textwidth]{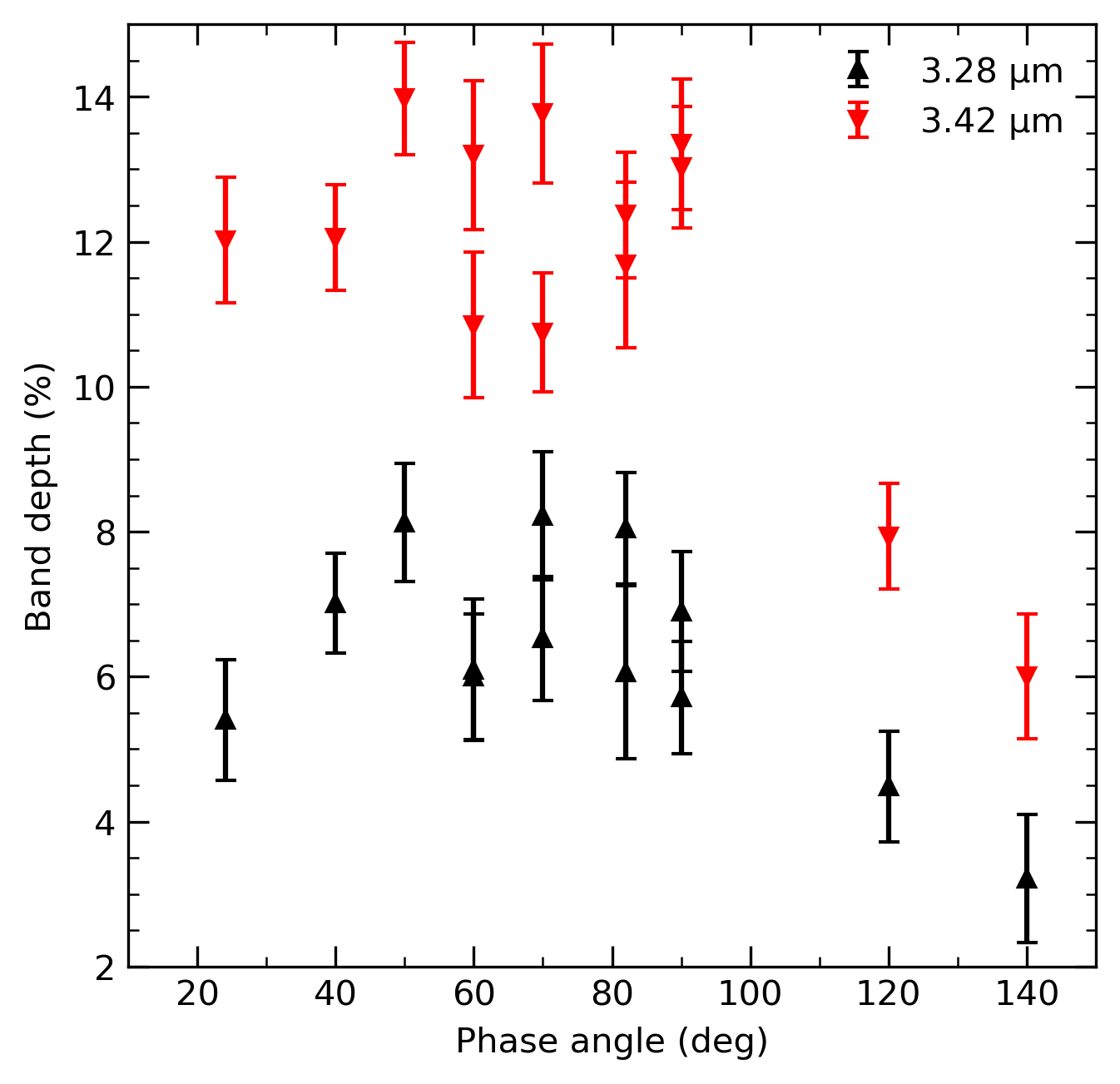}}
     \caption{Geometry observation effects on the 3.28 µm and 3.42 µm band depth for a mixture composed of PSD (70 vol.\%) and DECS-19 (30 vol.\%). Red triangles represent points for the 3.42 µm absorption band depth and black triangles represent points for the band depth of the 3.28 µm feature. Measurements for this mixture were made with incidence and emission angles ranging from 12$\degree$ to 70$\degree$. Points at a same phase angle correspond to various incidence and emission configuration.}
     \label{spectral_phase}
\end{figure}

We acquired 13 spectra of a mixture 70 vol.\% PSD, 30 vol.\% DECS-19 with phase angle from 24$\degree$ to 140$\degree$. For each spectrum, we determined the band depth of the 3.28 and 3.42 µm absorption bands to plot them according to the phase angle of the measurement. The result of this investigation is given in Fig. \ref{spectral_phase}. We can note the difference in the band depth between measurements in Figs. \ref{fig:SIM1_orga_det} and \ref{spectral_phase}. Different measurements were acquired on this mixture and the surface observed of the mixture is thus different between the nominal geometry configuration and the BRDF measurement. In Fig. \ref{spectral_phase}, we noticed a quite constant behavior of the band depth as a function of the phase angle, except at large phase angles ($>$ 100$\degree$) where the band depth is severely reduced. The band depth is also directly related to the mixture observed and to the SHADOWS reflectance measurements. The latter is strongly dependent on the surface observed by SHADOWS that varies even between measurements for the same phase angle. To avoid a decrease in the band contrast and therefore of the band depth, MIRS observations should avoid excessive phase angles. For a given phase angle, variations of the band depth parameter are quite small, generally lower than 4\%.\par

\begin{figure}
     \centering
     \resizebox{8cm}{!}{\includegraphics[width=\textwidth]{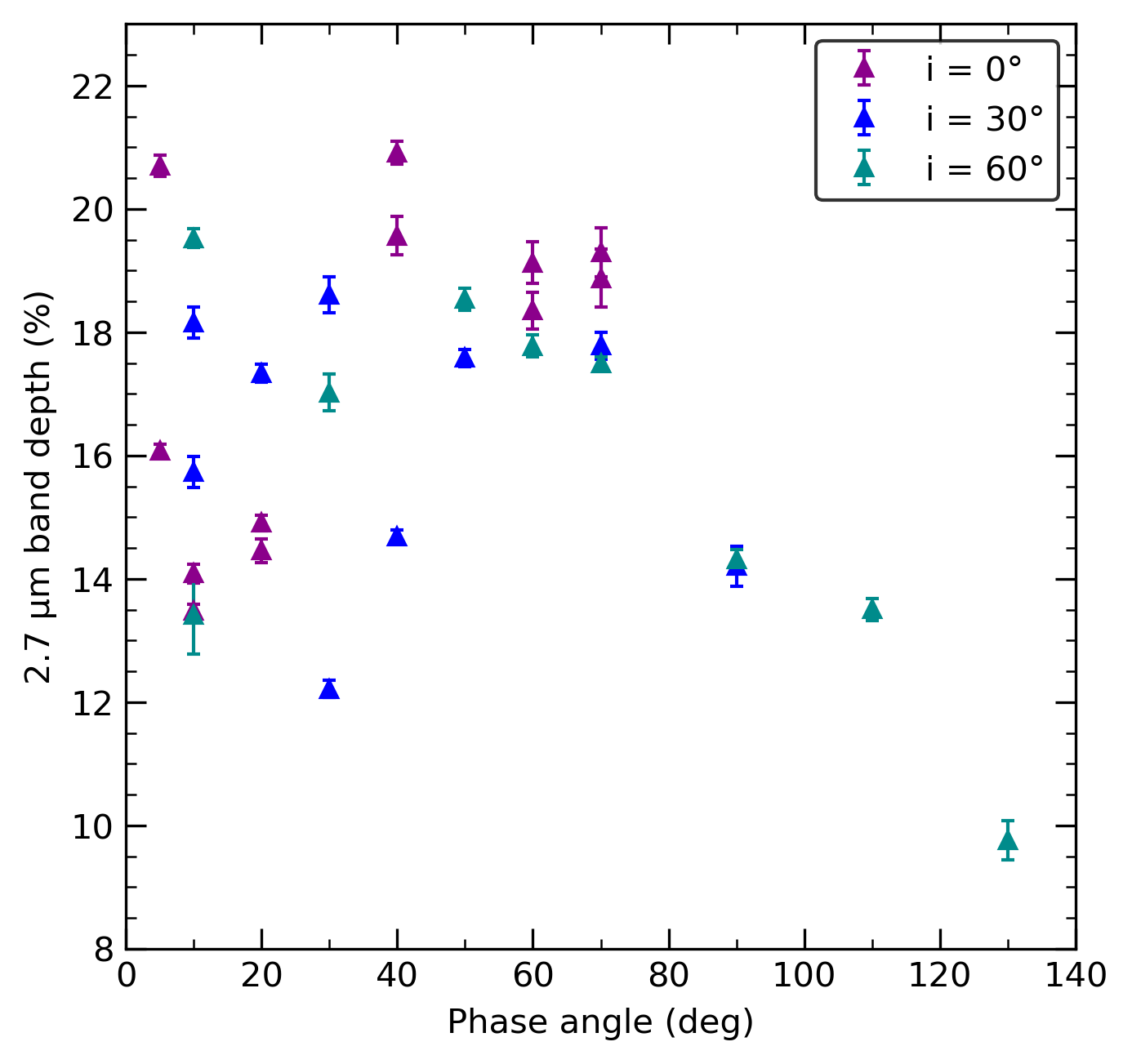}}
     \caption{Evolution of the 2.7 µm band due to hydrated minerals in the OPPS simulant with the phase angle and the incidence angle.}
     \label{hydmin_band_evolution_with_phase}
\end{figure}

\paragraph{Evolution of the hydrated minerals band's depth} \mbox{} \\
Similar to the investigation of organics, we analysed the evolution of the 2.7 µm band depth using BRDF measurements on the UTPS and OPPS simulants (Fig. \ref{hydmin_band_evolution_with_phase}). The band depth did not show a clear pattern in its evolution with the phase angle. However, it was observed that the minimum band was consistently computed for the larger phase angle (120-130°). This minimum demonstrates weak behaviour. It may only result in a 1\% reduction of the band depth or the feature may be significantly reduced by a factor of two compared to measurements at other phases. 
\par
We do not acquire data at phase angles lower than 5$\degree$ due to instrumental limitations. Our measurements were not affected by the opposition effect which generally occurs for phase angles less than 10$\degree$ \citep{Hapke_1986, Kitazato_2008, Fornasier_2015, Fornasier_2020}. For further investigation, the evolution of the bands at very small phase angles should be examined, as only a few measurements of opposition exist in the literature (e.g., \citealt{Nelson_2000}). \par
Some studies exist in the literature about variability of features with the geometry from laboratory measurements or from remote-sensing data. For example, the evolution of the 3 µm feature \citep{Pommerol_2008, Takir_2015, Potin_2019, Wargnier_2023_1}, the 0.6 µm absorption band \cite{Shepard_2011}, the 2.74 µm band \citep{Fornasier_2020} were investigated. All these works have shown the same effect we observed in this paper, namely that there are no significant variations in the band depth with phase angle and/or that decreases in the absorption band depth are possible for very high phase angles.

\subsubsection{Phase reddening}
We have computed the phase reddening coefficient $\gamma$ and the zero-parameter $Y_{0}$ to provide additional information for the interpretation of future Phobos observations. The OPPS exhibits $\gamma$ = 0.013 $\pm$ 0.002 10$^{-4}$ nm$^{-1}$/$\degree$ and $Y_{0}$ = 1.9 $\pm$ 0.1 \%.(100 nm)$^{-1}$. We found  $\gamma$ = 0.012 $\pm$ 0.002 10$^{-4}$ nm$^{-1}$/$\degree$ and $Y_{0}$ = 0.83 $\pm$ 0.09 \%.(100 nm)$^{-1}$ for the UTPS sample. These phase reddening parameters were computed based on a linear phase reddening assumption and from a spectral slope between 1.5 and 2.4 µm. For a discussion on phase reddening in the Phobos context, the reader is referred to \cite{Wargnier_2023_1, Wargnier_2023_2}. The phase reddening is a complex phenomenon and is generally attributed to the microscopic roughness (i.e., microns/sub-microns scale, \citealt{Beck_2012, Schroder_2014}) or increase of multiple scattering at large phase angle with the wavelength \citep{Hapke_2012c}. 

\subsubsection{Phase curve}
\begin{figure*}
\resizebox{\hsize}{!}{\includegraphics{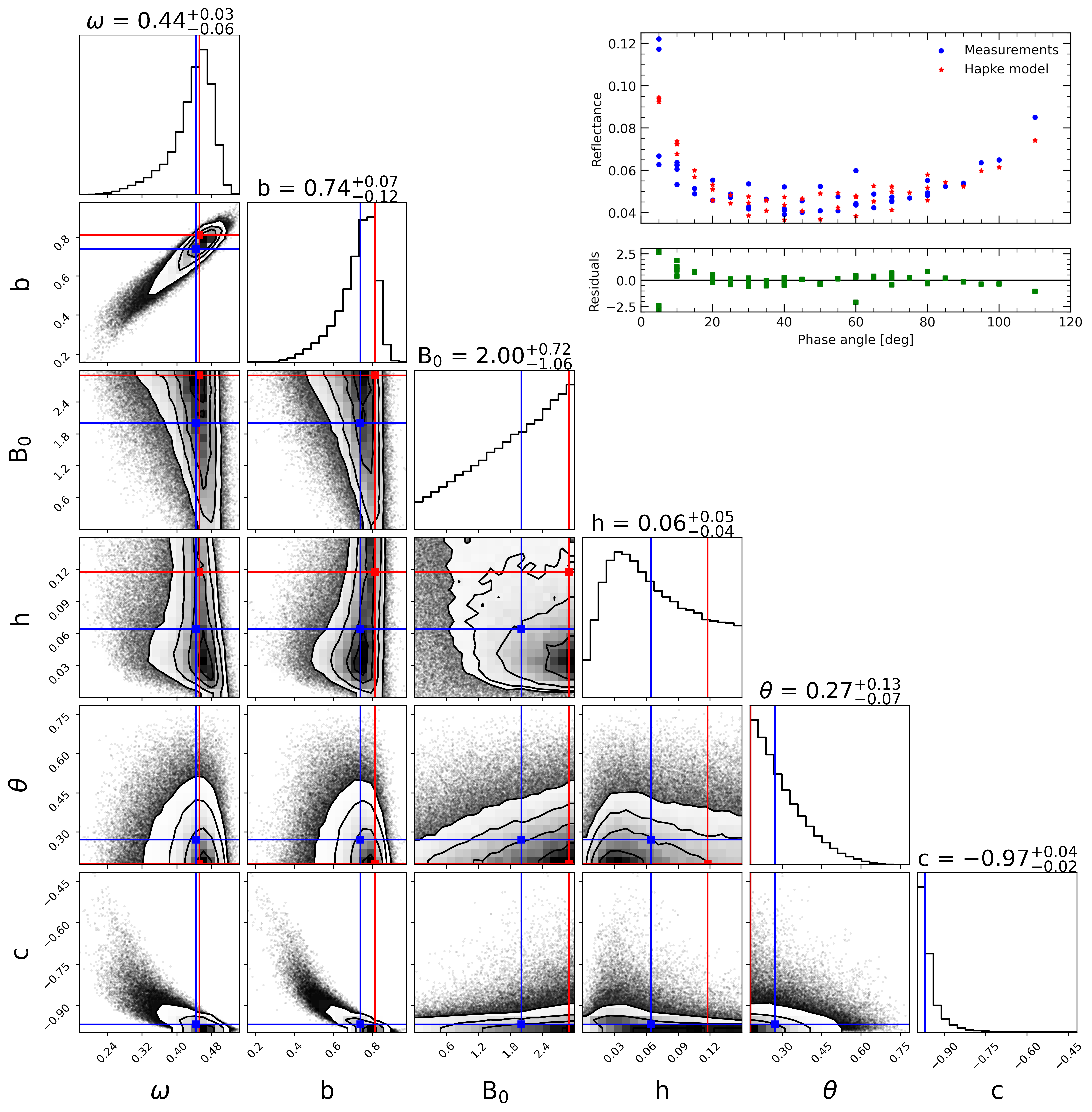}}
\caption{Posterior probability density function (PDF) of the Hapke parameters from the UTPS phase curve inversion. Each 1D histogram represents the 45000 accepted solutions. The 2D histograms show the correlation between the parameters. The blue line represents the median for a given parameter and the red line is the MLE. We chose to use the median of the parameter as the best-fit parameter. The values and the associated uncertainties are given as title of the 1D histogram. Top right: Hapke modelisation using the best-fit parameters found from the MCMC inversion, compared to the experimental UTPS phase curve. The residuals are also plotted, and given in percent.} 
\label{fig:mcmc_result_UTPS}
\end{figure*}

\begin{figure*}
\resizebox{\hsize}{!}{\includegraphics{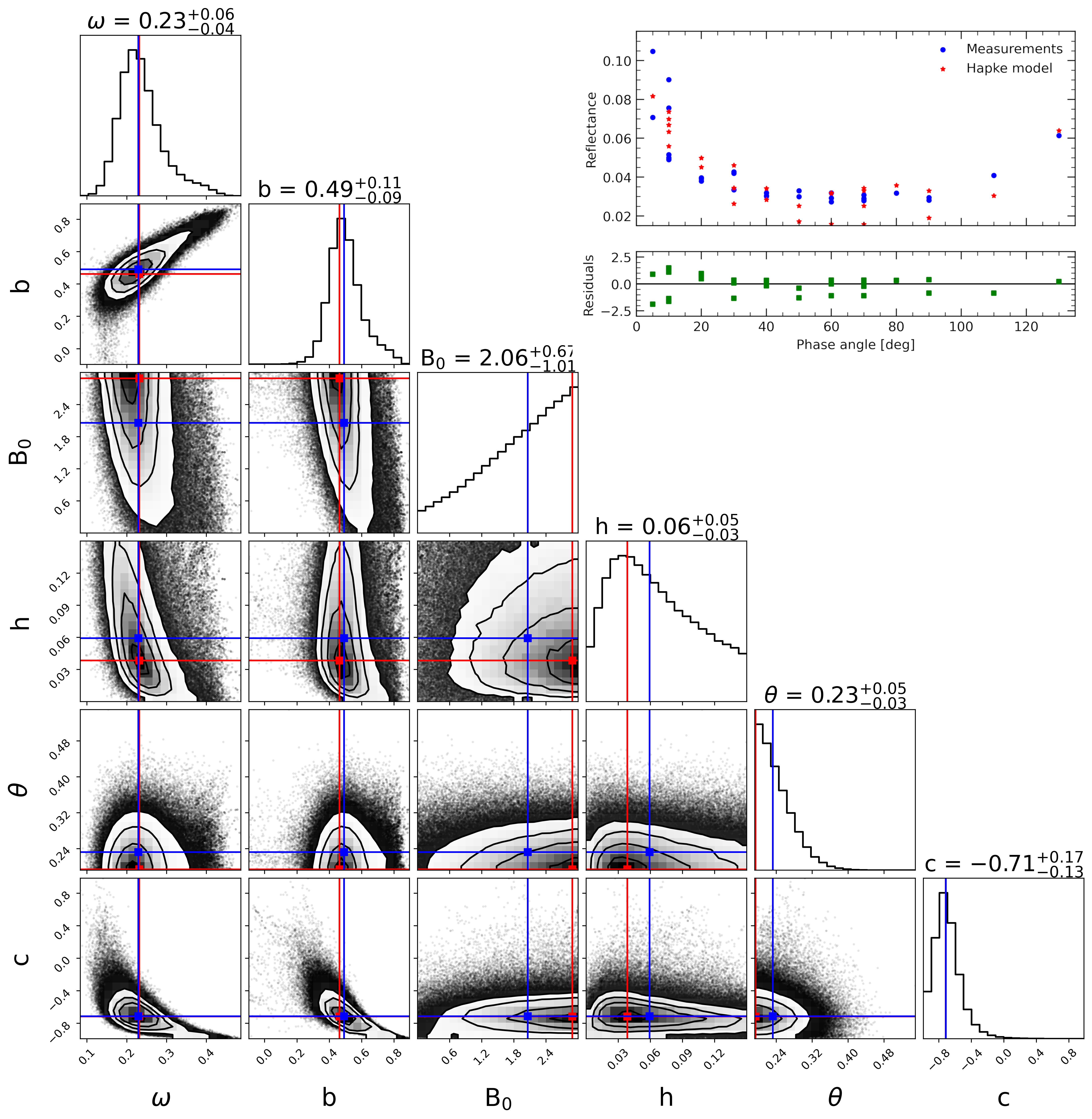}}
\caption{Posterior probability density function (PDF) of the Hapke parameters from the OPPS phase curve inversion. Each 1D histogram represents the 45000 accepted solutions. The 2D histograms show the correlation between the parameters. The blue line represents the median for a given parameter and the red line is the MLE. We chose to use the median of the parameter as the best-fit parameter. The values and the associated uncertainties are given as title of the 1D histogram. Top right: Hapke modelisation using the best-fit parameters found from the MCMC inversion, compared to the experimental OPPS phase curve. The residuals are also plotted, and given in percent.} 
\label{fig:mcmc_result_MIX98}
\end{figure*}

One of the main objective of this paper is to examine the evolution of the reflectance with the phase angle. The phase curve provides valuable insights into the surface, particularly its texture, and can provide crucial information for optimizing future observations of the JAXA/MMX mission, such as exposure time. The measurements indicate a decrease in reflectance from 0 to 70°, followed by an increase in reflectance at phase angles greater than 70-80° (Figs. \ref{fig:mcmc_result_UTPS} and \ref{fig:mcmc_result_MIX98}). The measurements appear to be quite scattered, in particular with a significant reflectance variability for the measurements at $\alpha$ = 5°. This variance is due to the fact that a measurement at a given phase angle is obtained with several combinations of incidence and emission angles. In particular, the incidence angle seems to have an extremely important effect at low phase angles \citep{Wargnier_2023_2}. In order to quantitatively compare with other laboratory measurements or with \textit{in situ} data, we modeled the phase curve of the Phobos simulants using the Hapke IMSA model \citep{Hapke_2012a}. A broad collection of incidence and emergence angles allows better constraining of Hapke parameters like the average roughness slope. \par
To avoid too many free parameters, we first set the single-lobe Henyey-Greenstein (1T-HG) function as single particle phase function into the IMSA model. However, this function fails to fit the measurements because of the strong forward-scattering that occurs at high phase angle. Therefore, we used the two-term Henyey-Greenstein (2T-HG) described in Sect. \ref{sect:hapke_model}. \par
Using MCMC inversion to fit the IMSA model to the experimental phase curves of the simulants, we were able to constrain the parameters (Figs \ref{fig:mcmc_result_UTPS} and \ref{fig:mcmc_result_MIX98}). The six Hapke parameters are then used to model the phase curve. The residuals of this fit are relatively low, except at $\alpha$ = 5° where the data variance lead to higher residuals. It appears very challenging to compare directly values obtained in laboratory and from remote-sensing data. The opposition effect especially appears to be stronger with \textit{in situ} data and the reflectance increases also extremely rapidly at large phase angles in laboratory indicated a strong forward-scattering (e.g., \citealt{Fornasier_2023}). However, we can more easily compare the obtained Hapke parameters with those associated to other laboratory experiments. Some studies are of particular interest in the case of Phobos. For example, \cite{Beck_2012} made a photometric studies of meteorites, \cite{Pommerol_2013} obtained the phase curve of the JSC Mars-1 analog, and \cite{Mandon_2021} measured the bi-directional reflectance distribution function (BRDF) of a martian meteorite (shergottite). However, \cite{Mandon_2021} does not apply any inversion technique to model parameters to their phase curve. We therefore decided to perform the inversion on this dataset using the same method as for our measurements, to allow quantitative comparison with the other studies. \newline
Table \ref{tab:Hapke_parameters} presents the derived Hapke parameters. The single-scattering albedo (SSA) of the UTPS and the OPPS show significant differences. The SSA value for UTPS falls between the values for the Allende meteorite and the JSC Mars-1 analog, but both are larger than the Tagish Lake meteorite SSA. The intensity of the opposition effect (B$_0$) is the same within the error bars for both simulants, but larger compared to other laboratory investigations such as Tagish Lake. The half-width of the shadow-hiding opposition effect $h_{sh}$ of the simulants appears to be larger than that of carbonaceous chondrites, but smaller than that of the JSC Mars soil analog and the NWA 4766 shergottite. The roughness is comparable for all samples, except for the lunar meteorite studied in \cite{Beck_2012}, which exhibits approximately twice the values of the others. The b and c parameters of the Phobos simulants indicate a predominantly forward-scattered surface for both, but with a higher and narrower lobe for the UTPS. The UTPS exhibits a scattering behaviour similar to that of the lunar meteorite studied in \citep{Beck_2012}, while OPPS is more similar to the Forest Vale meteorite \citep{Beck_2012}. Notably, the shergottite NWA4766 is particularly similar to the Tatahouine meteorite \citep{Beck_2012} in terms of both scattering and opposition effect properties (see Figs. 1 and 2). The JSC Mars-1 \citep{Pommerol_2013} analog exhibits different scattering properties compared to CCs, NWA 4766, and Phobos simulants, with a back-scattering broader lobe.

\begin{table*}
	\centering
	\caption{Hapke parameters derived from experimental phase curve at 0.6 µm measured with SHADOWS on different samples. The inversion was made using the MCMC method. The asymmetry parameter c is recomputed for the data of \cite{Beck_2012} and \cite{Pommerol_2013} because they used a different definition of the 2T-HG. $^{*}$g for 1T-HG and b for 2T-HG. $^{a}$For these samples, \cite{Beck_2012} used estimate uncertainties from \cite{Shepard_2007}. The filling factor was converted to the h$_{sh}$ parameter using the formula given in \cite{Helfenstein_2011}. $^{b}$The reader is referred to \cite{Pommerol_2013} for associated uncertainties of the Hapke parameters. The B$_{0}$ parameter was set with bounds [0,1]. $^{c}$ We used the BRDF data of NWA 4766, a basaltic shergottite, from \cite{Mandon_2021} to retrieve the Hapke parameters using the same inversion methods as our samples. It is important to note that the laboratory measurements from \cite{Beck_2012} and \cite{Mandon_2021} were performed on non-sieved powder but with an average size of 20 µm. $^{d}$Phobos Hapke parameters derived in \cite{Simonelli_1998} from Viking mission data acquired with a clear filter. $^{e}$Phobos Hapke parameters derived in \cite{Fornasier_2023} from the Mars Express HRSC/SRC dataset in the green filter.}
	\label{tab:Hapke_parameters}
    \resizebox{\textwidth}{!}{
	\begin{tabular}{ccccccccc} 
		\hline
		\textbf{Sample type} & \textbf{Sample} & \textbf{w} & \textbf{g or b$^{*}$} & \textbf{c} & \textbf{B$_{sh,0}$} & \textbf{h$_{sh}$} & \textbf{$\bar{\theta}$} & \textbf{References} \\
		\hline
		   \\[-1em]
		   Powder & UTPS & 0.44$^{+0.03}_{-0.06}$ & 0.74$^{+0.07}_{-0.12}$ & -0.97$^{+0.04}_{-0.02}$ & 2.00$^{+0.72}_{-1.06}$ & 0.06$^{+0.05}_{-0.04}$ & 15.46$^{+2.29}_{-1.15}$ & This work\\
		   \\[-1em]
		   & OPPS & 0.23$^{+0.06}_{-0.04}$ & 0.49$^{+0.11}_{-0.09}$ & -0.71$^{+0.17}_{-0.13}$ & 2.06$^{+0.67}_{-1.01}$ & 0.06$^{+0.05}_{-0.03}$ & 13.18$^{+2.86}_{-1.72}$ & This work\\
		   \\[-1em]
		   & Tagish Lake$^{a}$ & 0.157 & 0.431 & -0.436 & 0.334 & 0.023 & 14.1 & \cite{Beck_2012}\\
		   & Allende$^{a}$ & 0.399 & 0.366 & -0.318 & 0.720 & 0.020 & 12.8 & \cite{Beck_2012}\\
		   & Lunar Meteorite$^{a}$ & 0.850 & 0.746 & -0.956 & 0.843 & 0.000 & 25.7 & \cite{Beck_2012}\\
		   & JSC Mars-1$^{b}$ & 0.526 & 0.187 & 0.454 & 1.0 & 0.083 & 13.3 & \cite{Pommerol_2013}\\
		   & NWA 4766$^{c}$ & 0.94$^{+0.02}_{-0.02}$ & -0.03$^{+0.27}_{-0.24}$ & 0.15$^{+0.71}_{-0.99}$ & 1.85$^{+0.80}_{-1.01}$ & 0.11$^{+0.03}_{-0.05}$ & 24.64$^{+9.74}_{-6.87}$ & \cite{Mandon_2021} and this work\\
		   \\[-1em]
		\hline
        Phobos$^{d}$ & & 0.070 & -0.13 & N/A & 4$^{+6}_{-1}$ & 0.055 $\pm$ 0.025 & 22 $\pm$ 2 & \cite{Simonelli_1998}\\
        Phobos$^{e}$ & & 0.074 $\pm$ 0.002 & -0.301 $\pm$ 0.007 & N/A & 2.283 & 0.0573 & 24 & \cite{Fornasier_2023}\\
        \hline
	\end{tabular}
    }
\end{table*}

\section{Discussion}
\subsection{Single-scattering albedo}
As it can be noticed in Table \ref{tab:Hapke_parameters}, the single-scattering albedo $\omega$ (SSA) appears to be much higher with laboratory experiments, compared to the remote-sensing data. All dark samples including Tagish Lake, Allende have a SSA larger than 0.1. In particular, despite the very dark appearance of the simulants and the reflectance measurements (3-4\% at 600 nm), the SSA of these samples are 30 to 60 times higher than the value of the Phobos surface derived from photometric observations \citep{Simonelli_1998, Fornasier_2023}. The discrepancy between laboratory investigations and in-situ data can be attributed to two main factors. First, the grain size distribution is not the same in the laboratory as it is on the surface of small bodies. While the grain sizes used in this work are relatively small ($<$100 µm), asteroids generally have a wider grain size distribution (e.g., \citealt{Dellagiustina_2019, Michikami_2019, Burke_2021, Ogawa_2022}), sometimes with the presence of boulders (e.g., \citealt{Dellagiustina_2019, Michikami_2019}) that scatter light differently than small grains. This grain size effect could lead to the observation of a higher SSA. The other main reason is linked to the space environment in which asteroids evolve. Surface grains are exposed to space weathering, which modifies their spectroscopic, photometric, and chemical properties, and the low-gravity environment implies also that the surface is likely organized differently from the samples studied in the laboratory, with voids between particles, high micro-/macro-porosity, and high micro-/macro-roughness.\newline
It is also essential to note that the SSA is not really something observable, but rather a modelling parameter that describes the light behaviour on this grain. More precisely, it represents the albedo in the ideal case of an isolated single particle.

\subsection{Opposition effect}
\begin{figure}
 \centering
\resizebox{8cm}{!}{\includegraphics{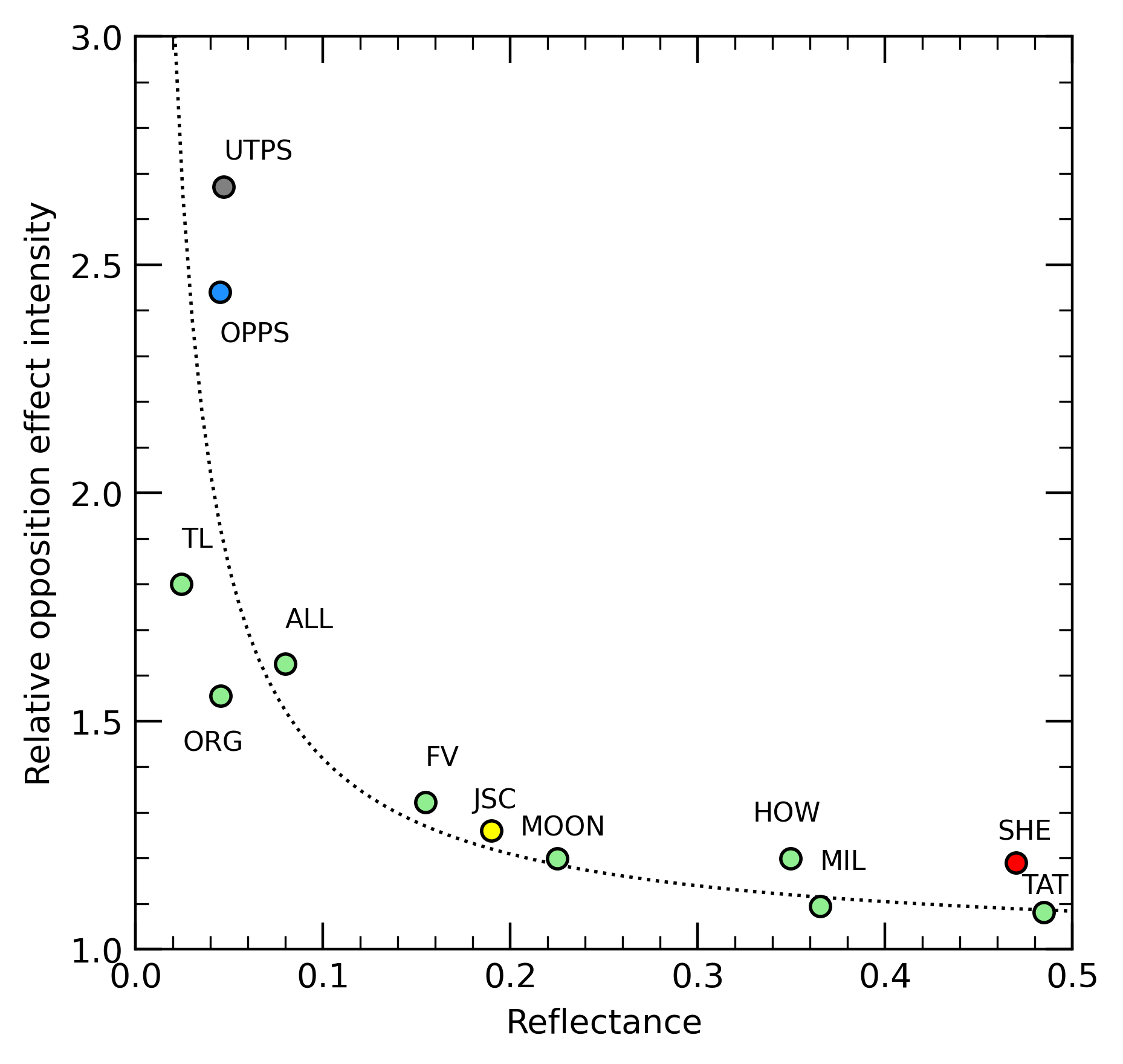}}
\caption{Figure adapted from \cite{Beck_2012}. Relative opposition effect intensity [ratio of the reflectance at $\alpha$ = 0° to that at $\alpha$ = 30° (i = 0°, e = 30°)] of laboratory samples: meteorites (green points) from \cite{Beck_2012}, JSC Mars-1 analog (yellow point) from \cite{Pommerol_2013} and a martian meteorite (basaltic shergottite, NWA 4766; red point) from \cite{Mandon_2021} and the two simulants of this study (gray and blue points). Note the small difference in the wavelength used in this work ($\lambda$ = 600 nm) and in \cite{Beck_2012} ($\lambda$ = 650 nm). The reader is referred to \cite{Beck_2012} for more informations about the samples (green points). Phobos ROIE has been found to be higher than the Phobos simulant, with a value of 3.6 \citep{Fornasier_2023}.} 
\label{fig:OE_lab_comp}
\end{figure}

The opposition effect is a relatively well-known effect that occurs in planetary surface and leads to a nonlinear increase of the reflectance at low phase angles (e.g., \citealt{Hapke_1998}). Despite the fact that laboratory measurements at a very low phase angle are rare, it is possible to extrapolate the SHOE trend for very small phase angles from measurements at $\alpha$ = 5$\degree$-10$\degree$ using photometric model. By applying the Hapke IMSA model, we have been able to obtain the relative opposition effect \citep{Beck_2012}, defined as the ratio of the reflectance at $\alpha$ = 0° to that at $\alpha$ = 30° (i = 0°, e = 30°). Fig. \ref{fig:OE_lab_comp} shows the relative opposition effect intensity (ROEI) as a function of the reflectance for the carbonaceous chondrites investigated in \citep{Beck_2012}, for the JSC Mars-1 analog \citep{Pommerol_2013}, for the NWA 4766 shergottite \citep{Mandon_2021}, and for the two Phobos simulants from this work. It appears that the simulants have an SSA similar to Tagish Lake, Orgueil, or Allende, but a stronger opposition effect. The ROIE of the simulants is closer than Tagish Lake to the value found for the Phobos surface from remote sensing data \citep{Fornasier_2023}. It is also important to note that the shergottite is far from the Phobos values in both SSA and ROIE. \newline
The data obtained in this study were insufficient to accurately determine the SHOE parameters of the Hapke model. However, they do provide some insight into the behavior at small phase angles. The CBOE could not be constrained due to laboratory limitations, so we assumed that the entire opposition surge was only due to SHOE. Our findings indicate that Phobos simulants contain opaque scatterers in the sample, with B$_{0,sh}$ > 1 \citep{Shepard_2007}. Future laboratory work should focus on studying the opposition effect, particularly the CBOE.

\subsection{Hapke model interpretation and limitations}
\begin{figure}
\centering
\resizebox{8cm}{!}{\includegraphics{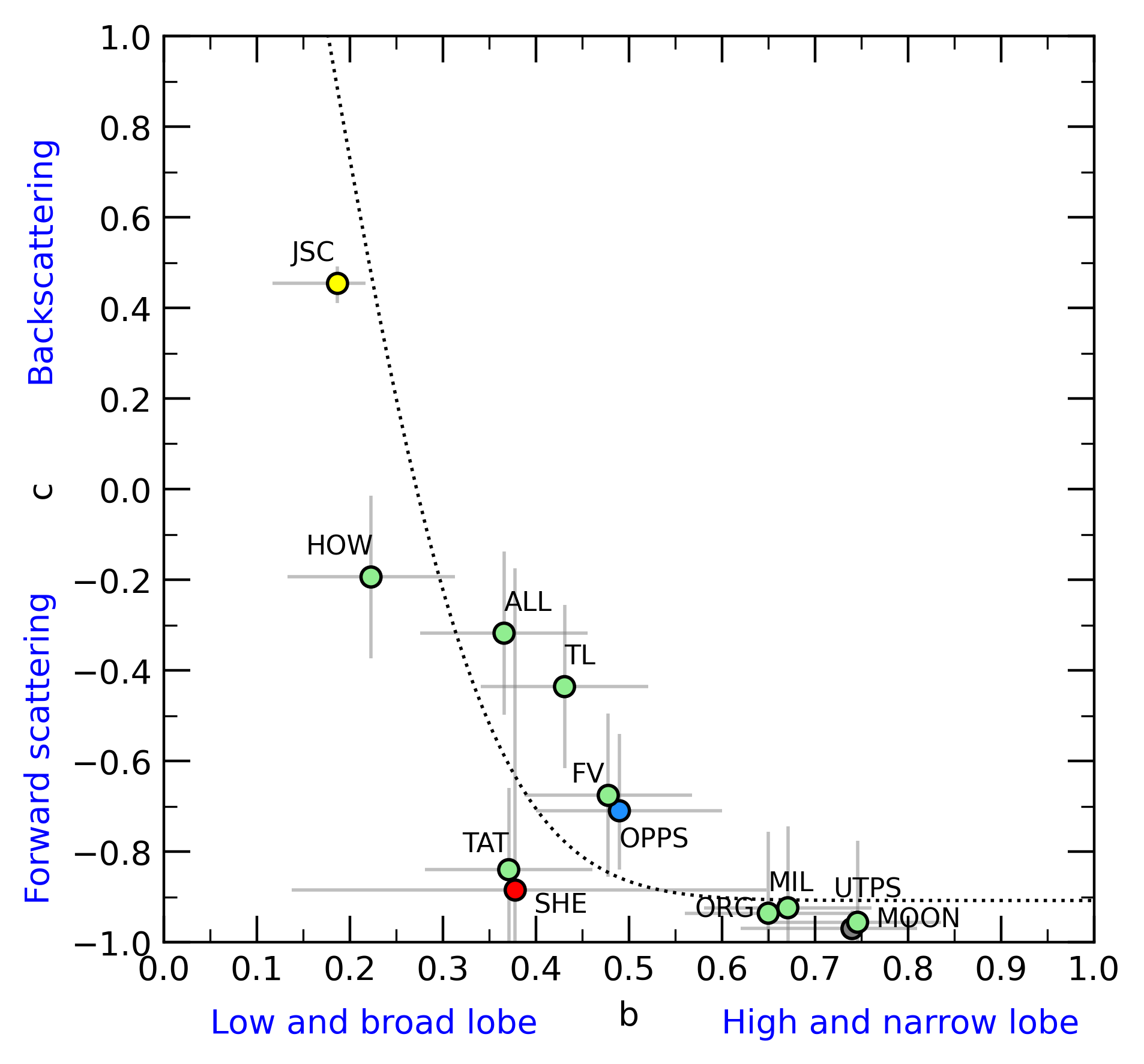}}
\caption{Asymmetry parameter c as a function of the shape parameter b. The dotted line represents the hockey stick relation \citep{Hapke_2012b}.} 
\label{fig:b_vs_c}
\end{figure}

The Hapke model is an interesting tool to try to better understand the surface of planetary bodies. It is particularly invaluable for the interpretation of spectroscopic and photometric data from remote-sensing measurements. Because modifications of the reflectance and light scattering could be due to several properties of a surface, and because of the nonlinearity and the relatively large number of free parameters, the global minimum of the model may be complicated to find. The model reflects the extremely complex light scattering process with intrinsic correlations between parameters and functions; like the asymmetric factor and the SSA, or the multiple scattering and the shadow-roughness functions. Therefore, retrieval of the physical properties of the samples from the derived parameters appears to be limited. A perfect BRDF data set should present data from 0 ° to 180 ° of phase angle with different incidence angle as suggested by \cite{Schmidt_2015}. High incidence measurements are particularly useful for constraining the photometric parameters, especially the surface roughness $\bar{\theta}$, as noted by \cite{Schmidt_2015} and \cite{Schmidt_2019}. However, it is extremely complicated to obtain such a dataset, and this study presents measurements between 5° and 130°. Then, we found some difficulties to well constrain the B$_0$, h, and $\bar{\theta}$ parameters. Data at smaller phase angle (<3°) are necessary for B$_0$ and at large phase angle (>120°) for $\bar{\theta}$. \newline
Additionally, it is well known that b and c are strongly dependent. We have observed a bimodal distribution for b and/or c in several test runs of the MCMC technique. This link between the two parameters is commonly referred to as the hockey-stick relation, as discussed in \cite{Hapke_2012b}, \cite{Fernando_2013}, and \cite{Schmidt_2015}. Fig. \ref{fig:b_vs_c} shows this effect with the laboratory measurements performed in \cite{Beck_2012}, in \cite{Pommerol_2013}, in \cite{Mandon_2021}, and in this work.\\

The two Phobos simulants in this study are spectroscopically and photometrically different from the NWA 4766 shergottite. However, they are closer to it than the JSC Mars-1 analog or other meteorites. The NWA 4766 shergottite is a typical basaltic material from Mars. Regarding the giant impact hypothesis, it would be worthwhile to investigate whether the spectroscopic and photometric behaviour of Phobos can be replicated by reproducing space weathering on a Martian meteorite, such as a shergottite, using ion irradiation.

\subsection{Implications for MMX observations of Phobos and Deimos}
\subsubsection{MIRS}
With its spectral range (0.9-3.6 µm), MIRS will be able to detect for the first time signatures of hydrated minerals and organics, if present on the surface of Phobos \citep{Barucci_2021}. Indeed, the presence of these signatures will give pivotal information about the origins of the two moons. \par
Observations of the Bennu surface for the OSIRIS-REx mission have shown that faint features with a few percent absorption band depth ($\sim$ 3\%) can be detected with a S/N $>$100 \citep{Simon_2020}. Hence, the MIRS spectrometer onboard MMX will be able to detect and characterize the 3.4 µm organics bands for more than 10 vol.\% (i.e., 4.51 wt.\% aromatic carbon and 0.87 wt.\% aliphatic carbon). In particular conditions, this threshold will be probably reduced. For 5 vol.\%, corresponding to 0.42 wt.\% aliphatic carbon and 2.21 wt.\% aromatic carbon, the detection will be more uncertain as the band depth is around 1\%, except for the 3.42 µm in PSD (see Fig. \ref{fig:SIM1_orga_det}). Such a feature can be complicated to detect by remote-sensing IR spectroscopy. \par
We compared the band depth and band center of our mixtures containing organics in various quantities with low-albedo asteroids. Values for the spectral parameters of these asteroids are given by \cite{Hromakina_2022}. Simulants with 10, 15, 20 vol.\% are close to asteroids of type B, C, P, or X. Of course, we probably don't expect large quantities of organic matter that cover the entire surface of Phobos, but some local variability is still possible. For example, for Bennu, \cite{Simon_2020b} have found C-H band depth varies from 1\% to more than 10\% in some precise locations. Hence, if an organic signature is present on Phobos, we can probably expect the 3.4 µm band depth to be included between 2 and 10\%. The expected value of the band depth at 3.4 µm will depend on the formation scenario for Phobos. In the hypothesis of the asteroid capture scenario, for a hydrous asteroid, the 3.4 µm C-H feature will exist with a maximum band depth of 10\% \citep{Takir_2019, Hromakina_2022}. For a partially dehydrated asteroid, the spectrum will present a 3.4 µm organic absorption band with a maximum depth of 5\% \citep{Takir_2019}. For a dehydrated asteroid, the existence of the 3.4 µm feature is more uncertain and it will be possibly linked with the presence of IOM (\citealt{Kaplan_2019}). In the others scenarios for the Phobos' origins and in particular, within the giant impact scenario, the C-H stretching modes of aliphatic and aromatic groups will be absent or only due to exogenous contribution \citep{DeSanctis_2019}. \newline
While the 30 vol.\% case of DECS-19 (i.e., $\sim$18 wt.\% of aliphatic and aromatic carbon) represents an extreme case with a deep organic band depth; values of 5 to 20 vol.\% appear to be quite realistic, with band depth values consistent with previous remote-sensing observations of organic-rich small bodies and with Ryugu sample band depth measurements by MicrOmega \citep{Pilorget_2022}. Direct measurements on Ryugu samples shows a organic content around 5\% \citep{Quirico_2023_2, Stroud_2024}. These results could help in understanding the organic content of Phobos materials from remote sensing measurements prior to sample return. \par
Hydrated minerals 2.7 µm signature are particularly persistent in the near-infrared. A very small quantity in a mixture with others silicates and/or opaque materials can make appear the feature \citep{Poggiali_2022, Poggiali_2023}. If present, a 2.7 µm O-H band depth from few percents to almost 40\% can be expected \citep{Takir_2019, Pilorget_2022} in the MIRS wavelength range considering a plausible quantity of phyllosilicate as expected in the case of the captured asteroid hypothesis. According to \cite{Fraeman_2014}, the possible 2.7 µm feature detected in CRISM data would have a band depth between 1 and 10\%, depending on the observed Phobos region. The 2.7 µm band for 1 vol.\% of phyllosilicates (i.e., 0.24 wt.\% OH groups) is still slightly visible but it is included in the noise. Hence, detectability for 1 vol.\% of phyllosilicates (band depth $\sim$ 3\%) will be more challenging. For 3 vol.\% of hydrated minerals (i.e., 0.73 wt.\% OH groups), the feature is already intense and reach a detectable band depth of 7\%. If present, this O-H feature will be detected by the MIRS spectrometer unambiguously for the first time  observing without gaps in the spectral range 0.9-3.6 µm, achieving the best spectral resolution for Phobos observations so far \citep{Barucci_2021}.\par
However, it is also interesting to note than phyllosilicate signatures including the 2.7 µm absorption band can be altered by space weathering processes \citep{Noguchi_2023, LePivert_2023}. Also, it has been shown that remote sensing observation of Ryugu from Hayabusa2 shows a two times shallower 2.7 µm band depth compared to measurement on the returned sample \citep{Pilorget_2022}. It is important to consider that this effect may not be systematic and could be challenging to quantify on other bodies. However, it is important to bear this in mind when comparing orbital data with laboratory data. \par
The detectability of both organics and hydrated minerals could strongly depend on several parameters including grain size, relative grain size between the endmembers, porosity, mixing, and composition. Therefore, the detectability limits derived from this work are not an absolute result but give indications for the future observations of the Phobos surface. This will be particularly helpful to interpret the remote-sensing observations of MIRS and quantify the Phobos surface composition before the analysis of the sample return scheduled in 2031. \par
As demonstrated in this study, obtaining measurements at various geometries can also provide valuable additional information about the surface. However, the exact interpretation can be debated as models have difficulties explaining all physical effects that occur in complex environments such as planetary regolith. Measuring the BRDF of return samples from Ryugu, Bennu, and/or Phobos could be an interesting endeavour. This would enable a comparison of BRDF measurements in both laboratory and remote-sensing contexts from the same body, providing a better understanding of laboratory experiment results. 

\subsubsection{TENGOO/OROCHI}
Considering the wavelength range of the measurements performed on our dataset, our results can be useful not only for interpretation of MIRS data but also in support for the OROCHI observations. TENGOO and OROCHI are the cameras of the MMX spacecraft \citep{Kameda_2021}. In particular, OROCHI will observe Phobos with 7 different filters to obtain UV-VIS spectrophotometry: 390 $\pm$ 50 nm, 480 $\pm$ 30 nm, 550 $\pm$ 30 nm, 650 $\pm$ 40 nm, 730 $\pm$ 40 nm, 860 $\pm$ 40 nm, 950 $\pm$ 60 nm. The 390 nm filter is unfortunately not cover by our measurements. For the other filters, we converted our spectra in OROCHI spectrum by considering rectangular filters (S. Kameda, personal communication). \par
\begin{figure*}
     \centering
     \begin{subfigure}[b]{0.49\textwidth}
         \centering
         \resizebox{\hsize}{!}{\includegraphics[width=\textwidth]{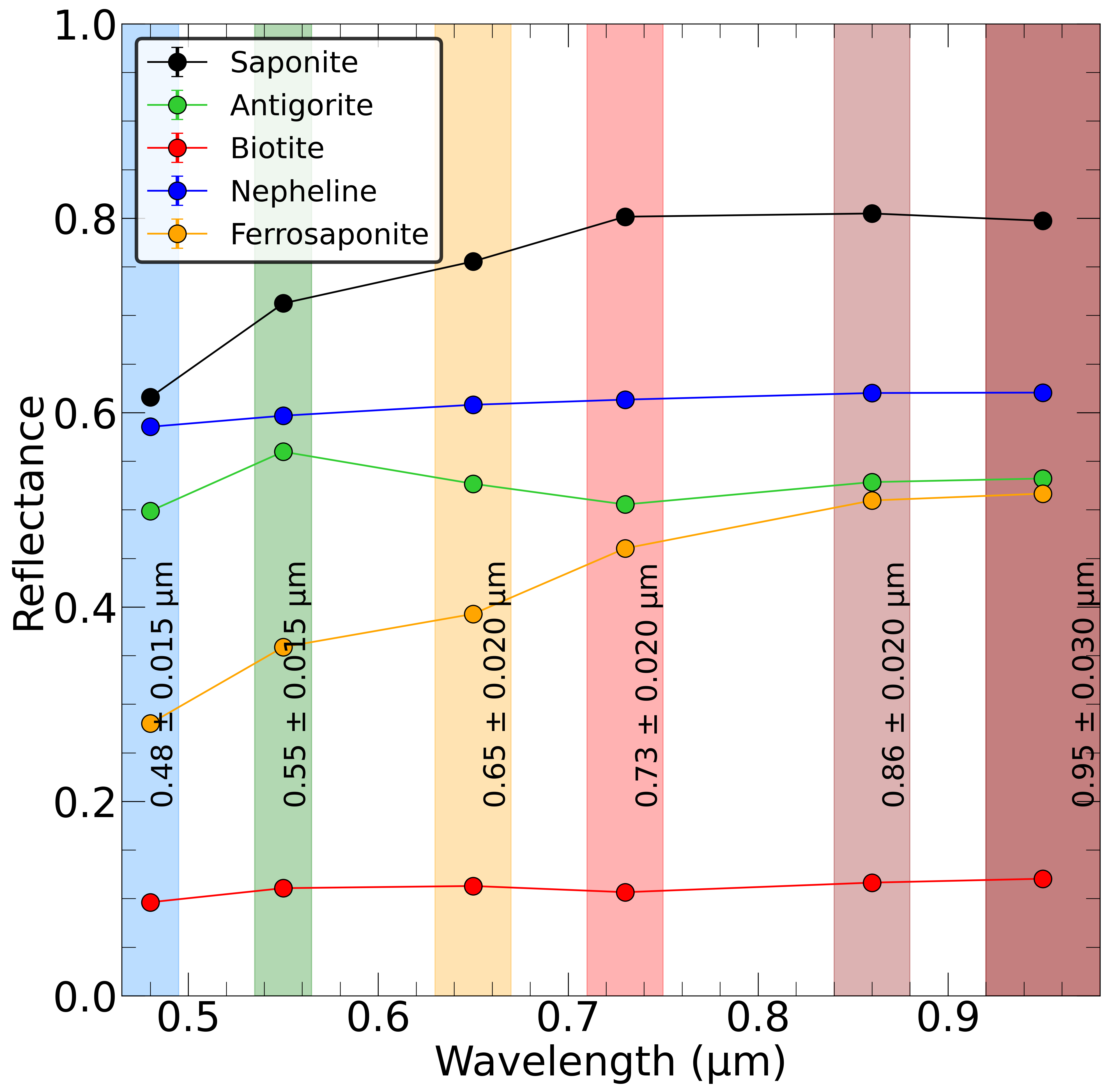}}
         \caption{}
         \label{fig:orochi_spec_endmembers}
     \end{subfigure}
     \hfill
     \begin{subfigure}[b]{0.49\textwidth}
         \centering
         \resizebox{\hsize}{!}{\includegraphics[width=\textwidth]{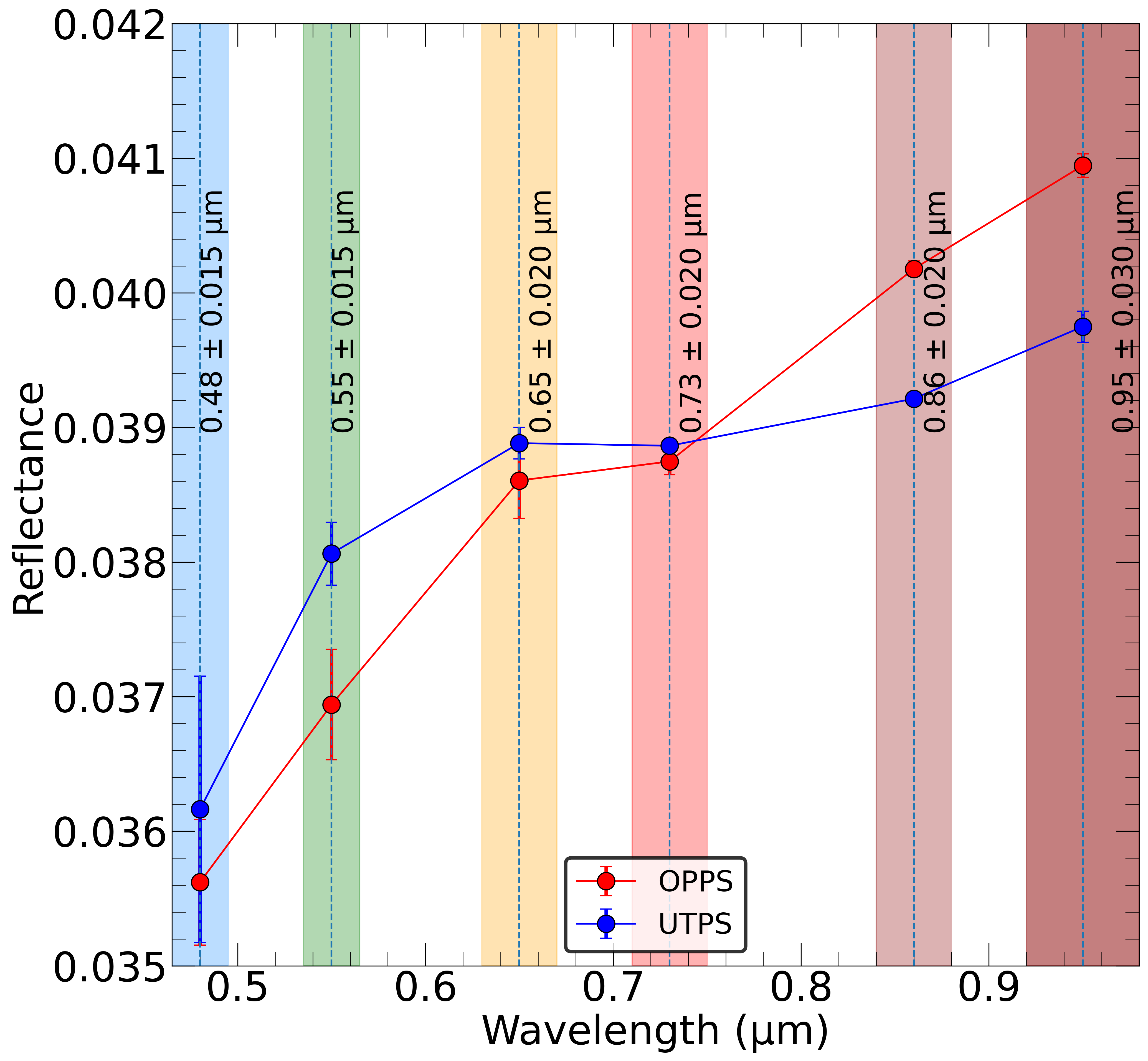}}
         \caption{}
         \label{fig:orochi_spec_simulant}
     \end{subfigure}
     \caption{Converted OROCHI spectra obtained from SHADOWS and SHINE measurements of phyllosilicates (50-100 µm). The OROCHI spectra were obtained assuming rectangular filters. The ferrosaponite (0-125 µm) spectrum was obtained from the RELAB database.}
     \label{fig:orochi_spec}
\end{figure*}

\cite{Fraeman_2014} found a possible 0.65 µm feature possibly linked to the presence of Fe-bearing phyllosilicate in the red unit. This band has been recently confirmed by the analysis of the NOMAD data \citep{Mason_2023}. OROCHI will investigate this broad and shallow feature. Our spectra (Fig. \ref{fig:orochi_spec_endmembers}) show that saponite is the only phyllosilicate of this study that exhibits such feature. Despite this band being really reduced because of a poor iron saponite, is it still visible and we computed a band depth (using a continuum defined as a straight line between the points at 0.55 µm and 0.73 µm) of $\sim$1\%. To compare with a Fe-rich phyllosilicate, we used a ferrosaponite (0-125 µm) spectrum of the RELAB database (id:c1jb762a). As expected, the band depth is more important in this case ($\sim$5\%). However, when looking at the OROCHI OPPS resampled spectrum (Fig. \ref{fig:orochi_spec_simulant}) the 0.65\% absorption band is not visible and probably removed by the opaques. This implies that, to have a visible 0.65 µm band, the Phobos surface should contains Fe-phyllosilicate particularly rich in Fe, and probably associated to less opaque materials than the OPPS. The CRISM data analysed by \cite{Fraeman_2014} reveal 0.65 µm-band depths  between 0.5 and 5\%. Therefore, if the feature is detectable by OROCHI, it may implies that the investigated part of the surface is particularly rich (and possibly >40-50\%) in iron-bearing phyllosilicate such as ferrosaponite. In this case, such an observation would favor the formation of Phobos and Deimos by a giant impact. Others minerals could play a role in this feature such as nontronite or cronstedtite \citep{Fraeman_2014}, but the band might be due to nanophase iron particles \citep{Fraeman_2014}. Such iron particles can be created by space weathering process \citep{Pieters_2000} such as ion irradiation \citep{Hapke_2001_SS, Brunetto_2005} or temperature alteration \citep{Lasue_2022}. \par
The OROCHI spectra will also allow to observe the visible spectral slope of Phobos. In particular, the data at 0.48 µm and 0.86 µm will be used for this purpose. The UTPS shows a spectral slope (0.48-0.86 µm) of 2.22 $\pm$ 0.06 \%/100nm, whereas the OPPS exhibits a higher slope of 3.37 $\pm$ 0.05 \%/100nm. The visible slope obtained from OROCHI observations could provide additional information that may aid in the interpretation of Phobos' composition.

\section{Conclusion}
We presented an experimental work on Phobos simulants. This work was, first, dedicated to a development of a Phobos simulant, for laboratory studies of physical change of the surface and for evaluation of organics and hydrated minerals detectability. This experimental study of a Phobos simulant led to several main results:
\begin{itemize}
    \item According to spectral slope, reflectance in the VNIR and to positions of the Christiansen features, the restrahlen band, and the transparency feature in the mid-infrared, a mixture of different (OPPS) endmembers composed of silicates, organics, and opaque materials, was found to be consistent with the Phobos spectra. One of the best match was obtained with a mixture made of olivine (20 vol.\%, 50-100 µm), anthracite (20 vol.\%, $<$1 µm), the DECS-19 coal (20 vol.\%, 50-100 µm), and saponite (40 vol.\%, 50-100 µm).
    \item We studied the detectability of CH-rich organic that could be found on Phobos in the case of captured asteroid hypothesis. MIRS will allow for the first time to observe Phobos in the CH stretching mode regions, characteristic of organic compounds usually found on small bodies. In this context, we have explored the feasibility of CH-rich organic detection in a Phobos simulant mostly composed of silicates and coal. Our results show that organics are detectable for 5.4 wt.\% of CH$_x$ in the simulant. If Phobos is indeed linked to a D-type captured asteroid, organic matter is expected \citep{Jones_1990, Brown_2016, Barucci_2018}. However, detectability will be probably strongly dependent on the composition of dark materials, grain size, organic source, etc. A definitive conclusion about the detectability is challenging, but anyway it is that these bands at 3.4 µm are weak and that it will be necessary to maximize the chances of seeing it with MIRS by observing Phobos at a low phase angle and with a maximum exposure time in order to increase the S/N.
    \item We also investigated the detectability of hydrated minerals. Up to now, the 2.7 µm O-H feature was not unambiguously detected on Phobos, but, if present, MIRS will be able to detect this feature. Using the same Phobos simulant of the organic study, we investigated the detectability of this absorption band. Our results show that the 2.7 µm O-H feature is intense and will be probably detectable for MIRS for more than 0.24 wt.\% of OH groups of the phyllosilicates. These quantities of hydrated minerals could be expected on Phobos \citep{Rivkin_2002, Fraeman_2014, Takir_2019}. 
    \item Our work shows that the inversion of the Hapke equation is a strenuous task as the parameters are correlated. Using Monte-Carlo Markov Chain bayesian inference, we were able to explore the parameters space and obtained the best-fit parameters of the experimental data. Some parameters may be difficult to constrain with laboratory data (B$_{0}$, h$_{sh}$, and $\theta$), while other parameters can be determined unambiguously ($\omega$, b, c). The SSA of the Phobos simulants falls within the same range as that of some carbonaceous chondrites, such as Tagish Lake and Allende. The opposition effect of the simulants seems to be higher than that of any other meteorites used for comparison.
\end{itemize}
This work was made in the context of the MMX mission and of the future MIRS observation of the martians moons. It aimed at better understanding spectro-photometric properties of Phobos laboratory simulants for the interpretation of the MIRS and TENGOO/OROCHI data. In the next papers, we will try to explore, in laboratory using the Phobos simulants, different effects that are present or occur on a surface of an airless body such as Phobos. In particular, porosity/roughness at the micro- and macro-scale, and space weathering are of interest for the surface of small bodies because spectroscopic and photometric remote sensing observations are affected by these parameters. However, these effects are non-exhaustive and further investigations could be dedicated to the study of other samples, e.g. more linked to a basaltic mars composition. We believe that the study of variation of geometry effects such as phase reddening and phase curve variations for different types of surfaces may be useful for other spectro-photometric observations of small bodies.

\section*{Acknowledgements}
This work was carried out in support for the MIRS instrument onboard the future MMX mission, with the financial support of the Centre National d’Etudes Spatiales (CNES). T.G and A.W thank the support from Agence National de la Recherche under the grant ANR-20-CE49-0004-01. We gratefully acknowledge iUMTEK for the LIBS analysis of the biotite sample. A. W. thanks Jamila El Bekri (LGPM) for technical support during the preparation of the samples and Cyril Breton (LGPM) for the measurements of grain size by laser diffraction method. We thank also Olivier Brissaud (IPAG) for the help with the SHADOWS and SHINE instruments. We are grateful to the La Mûre Museum (France) and the Université Grenoble Alpes (UGA, France) for donating the anthracite sample. We also acknowledge the Muséum National d'Histoire Naturelle (MNHN, France) for providing us antigorite (12 M.) and nepheline (MNHN-MIN-2011-4569) samples. This research utilizes a ferrosaponite (0-125 µm) spectrum acquired by Janice L. Bishop (id:c1jb762a) with the NASA RELAB facility at Brown University.

\section*{Data availability}
All spectroscopic data will be  available for further studies in the Zenodo repository (doi: 10.5281/zenodo.11087841). 

\appendix
\counterwithin{figure}{section}
\counterwithin{table}{section}

\section{Samples characterization: SEM images, EDX, and grain size distribution}
\begin{figure}
\centering
\resizebox{10cm}{!}{\includegraphics{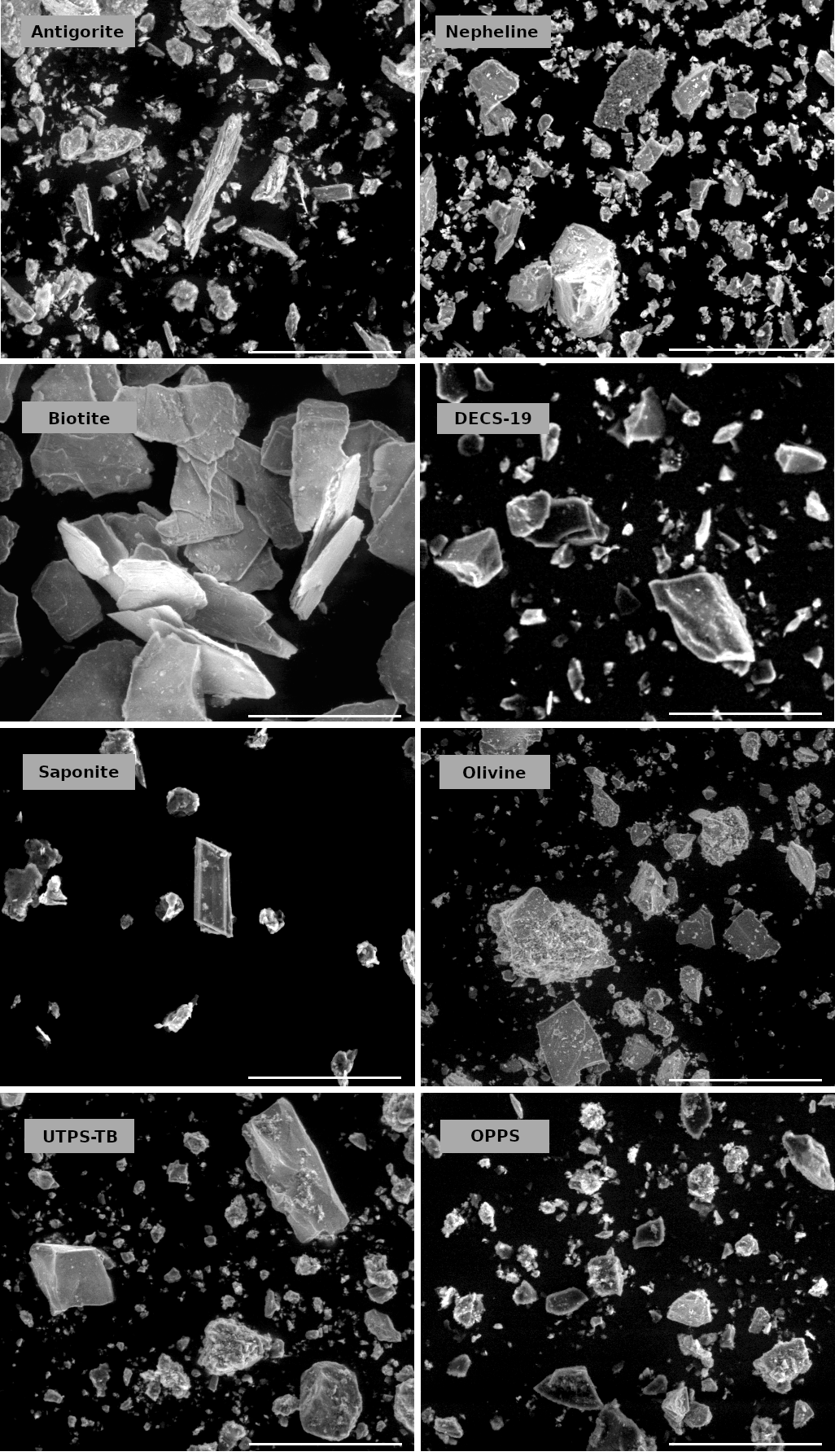}}
\caption{SEM images of some endmembers and of the two Phobos simulants used in this work. Note that the scale bar is the same for all SEM images and represents 100 µm.}
\label{fig:SEM_images}
\end{figure}

\begin{figure}
\centering
\resizebox{8cm}{!}{\includegraphics{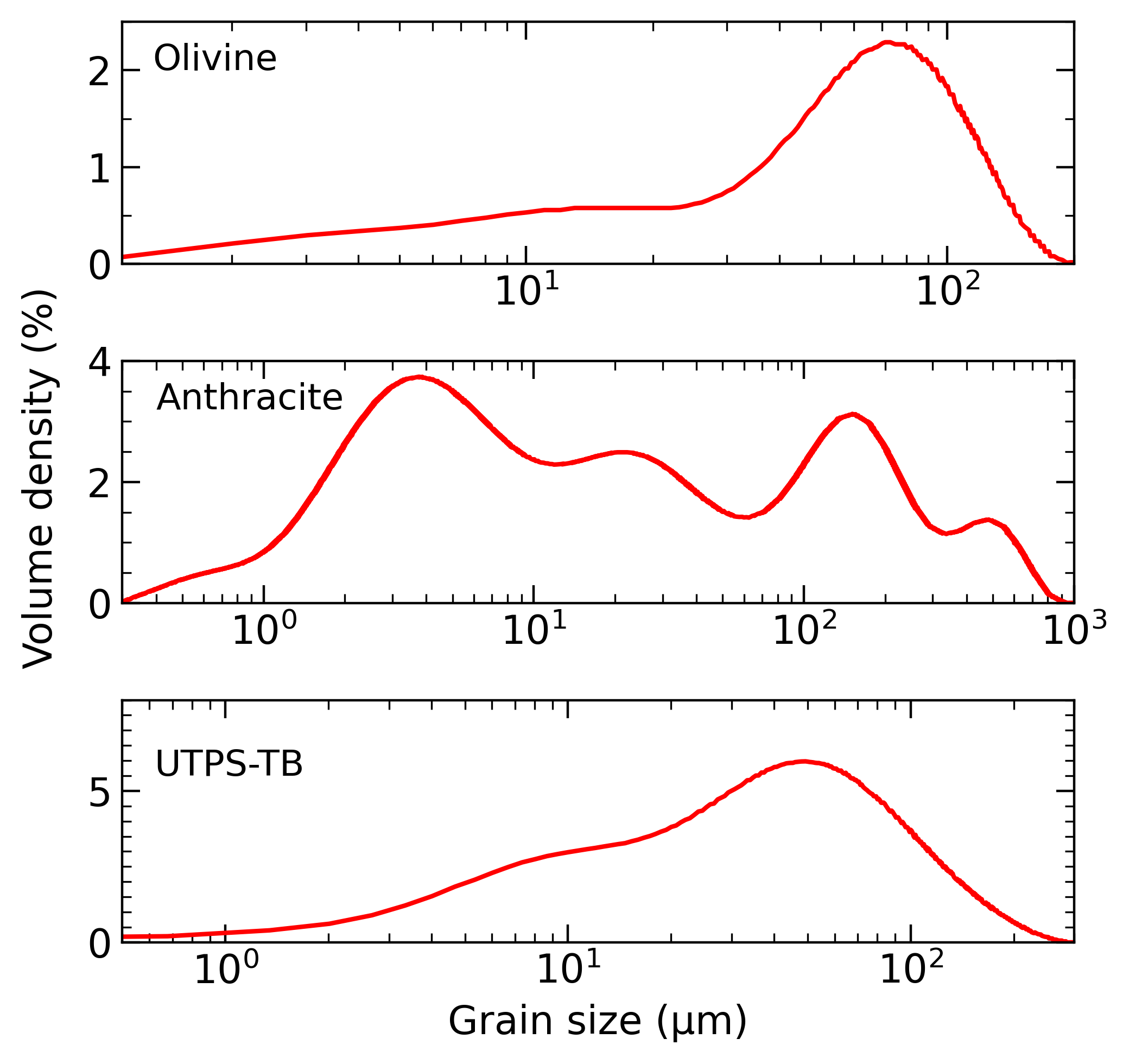}}
\caption{Granulometry of some samples used in this study from laser diffraction measurements.}
\label{fig:granulo}
\end{figure}

The following presents the results of the sample characterization performed, including SEM images of the mixtures and individual components (Fig. \ref{fig:SEM_images}), grain size distribution (Fig. \ref{fig:granulo}), characteristics of the endmembers (Table \ref{table:endmembers}), EDX elementary analysis (Table \ref{table:EDX}), and bulk chemical composition of the OPPS simulant (Table \ref{table:bulk_chemical_compo}).

\begin{table*}
\caption{End members' characteristics and properties. \label{table:endmembers}}
\centering
\begin{tabular}{cccccc}
\hline
\hline
Compound & Grain size & Density (g.$cm^{-3}$) & Density references & Origins \\ 
\hline
\textbf{Nesosilicates} & & & & \\
Olivine (Mg-rich) & 50-100 µm &  $\sim$ 3.3 & \cite{Sultana_2023} & Donghai Pellocci Crystal Products\\
\textbf{Phyllosilicates} & & & & & \\
Biotite & 50-100 µm &  $\sim$ 2.8 & Kremer Pigmente & Canada\\
Antigorite & 50-100 µm &  $\sim$ 2.6 & \cite{Mineralogy_1990} & Haute-Garonne (France)\\
Saponite SapCa-2 & 50-100 µm &  $\sim$ 2.3 & \cite{Mineralogy_1990} & California (USA)\\
\textbf{Feldspathoid} & & & & \\
Nepheline (eleolite) & 50-100 µm &  $\sim$ 2.6 & \cite{Mineralogy_1990} & Brevig (Norway)\\ \hline
\textbf{Organic compounds} & & & & \\
\begin{tabular}{c} Titan tholins \\ (95\% N$_2$ : 5\% CH$_4$) \end{tabular} & $\sim$ 400 nm & $\sim 1$ & \cite{Carrasco_2009} & LATMOS (PAMPRE) \\
DECS-19 & 50-100 µm & 1.75 & \cite{Vinogradoff_2021} & Penn State Coal Sample Bank \\ \hline
\textbf{Opaque materials} & & & & & \\
Anthracite & $<$ 1 µm & 1.62 & \cite{Sultana_2023} & La Mûre museum (France) \\
\hline
\end{tabular}
\end{table*}

\begin{table}
\caption{Elemental composition of the OPPS given by EDX analysis. Minor elements ($<$ 0.1 \%) and light elements are not reported. \label{table:bulk_chemical_compo}}
\centering
\begin{tabular}{cc}
\hline
\hline
Elements & Atomic (\%)  \\ 
\hline
C & 39.4 \\
Na & 2.2 \\
Mg & 24.9 \\
Al & 2.3 \\
Si & 24.9 \\
K & 0.1 \\
Ca & 4.7 \\
Fe & 1.5 \\
\hline
\end{tabular}
\end{table}

\begin{table}
\caption{EDX analysis of principal endmembers used in this work. \label{table:EDX}}
\centering
\begin{tabular}{cccc}
\hline
\hline
Minerals & Elements & Atomic (\%) & Weight (\%) \\ 
\hline
Antigorite & Mg & 19.5 &  23.2\\
           & Si & 15.8 &  21.7\\
           & Fe & 2.3 &  6.2\\
\hline
Augite & Na & 1.2 &  1.3\\
       & Mg & 8.1 &  9.2\\
       & Al & 4.4 &  5.5\\
       & Si & 19.2 &  25.1\\
       & Ca & 5.6 &  10.4\\
       & Ti & 0.1 &  0.2\\
       & Fe & 1.4 &  3.6\\
\hline
Biotite & Na & 0.7 &  0.7\\
        & Mg & 8.7 &  9.1\\
        & Al & 5.7 &  6.6\\
        & Si & 16.9 &  20.5\\
        & K & 4.8 &  8.1\\
        & Ti & 0.7 &  1.5\\
        & Fe & 5.9 &  14.2\\
\hline
Nepheline & Na & 8.7 &  9.7\\
          & Al & 12.9 &  16.8\\
          & Si & 17.2 &  23.4\\
          & K & 2.1 &  4.0\\
\hline
Olivine & Mg & 26.3 &  30.3\\
        & Si & 14.8 &  19.6\\
        & Fe & 3.0 &  7.7\\
\hline
Saponite & Na & 1.8 &  2.1\\
         & Mg & 10.8 &  13.2\\
         & Al & 1.8 &  2.4\\
         & Si & 15.1 &  21.1\\
         & Ca & 3.8 &  7.7\\
         & K & 0.1 &  0.3\\
\hline
\end{tabular}
\end{table}

\section{Comparison of the different Hapke's inversion methods}
This appendix presents the results of the various inversion methods of the Hapke equation explored in this study. The values of each parameter for each method and for UTPS and OPPS simulants are shown in Table \ref{tab:Hapke_inv_comparison}.

\begin{table*}
	\centering
	\caption{Comparison of the best-fit Hapke parameters obtained after the various inversion methods investigated. The CBOE effect was not considered. The italic values indicated that the parameters is at one of the bounds set for the inversion.}
	\label{tab:Hapke_inv_comparison}
    \resizebox{\textwidth}{!}{
	\begin{tabular}{ccccccccc} 
		\hline
		\textbf{Methods} & \textbf{Sample} & \textbf{w} & \textbf{b} & \textbf{c} & \textbf{B$_{sh,0}$} & \textbf{h$_{sh}$} & \textbf{$\bar{\theta}$} & \textbf{RMS}\\
		\hline
		   LM & UTPS & 0.47 $\pm$ 0.02 & 0.80 $\pm$ 0.05 & -0.98 $\pm$ 0.01 & \textit{3.0 $\pm$ 2.3} & \textit{0.15 $\pm$ 0.21}& \textit{11 $\pm$ 29} & 0.0095\\
		   & OPPS & 0.23 $\pm$ 0.15 & 0.46 $\pm$ 0.17 & -0.70 $\pm$ 0.33 & \textit{3.0 $\pm$ 3.3} & 0.036 $\pm$ 0.17 & \textit{11 $\pm$ 19} & 0.0109\\
		   \hline
		   Basin-hopping & UTPS & 0.41 $\pm$ 0.04 & 0.85 $\pm$ 0.04 & -0.995 $\pm$ 0.003 & \textit{3.0 $\pm$ 0.01} & 0.28 $\pm$ 0.18 & \textit{11 $\pm$ 0.01} & 0.0095\\
		   & OPPS & 0.23 $\pm$ 0.05 & 0.46 $\pm$ 0.08 & -0.71 $\pm$ 0.14 & \textit{3.0 $\pm$ 0.01} & 0.036 $\pm$ 0.04 & \textit{11 $\pm$ 0.01} & 0.0109\\ \hline
		   MCMC & UTPS & 0.44$^{+0.03}_{-0.06}$ & 0.74$^{+0.07}_{-0.12}$ & -0.97$^{+0.04}_{-0.02}$ & 2.00$^{+0.72}_{-1.06}$ & 0.06$^{+0.05}_{-0.04}$ & 15.46$^{+2.29}_{-1.15}$ &  0.0101\\
		   & & & & & & & & \\
		   & OPPS & 0.23$^{+0.06}_{-0.04}$ & 0.49$^{+0.11}_{-0.09}$ & -0.71$^{+0.17}_{-0.13}$ & 2.06$^{+0.67}_{-1.01}$ & 0.06$^{+0.05}_{-0.03}$ & 13.18$^{+2.86}_{-1.72}$ & 0.0126\\
        \hline
	\end{tabular}
    }
\end{table*}

\section{Vacuum effect for measurements}
\begin{figure}
\centering
\resizebox{8cm}{!}{\includegraphics{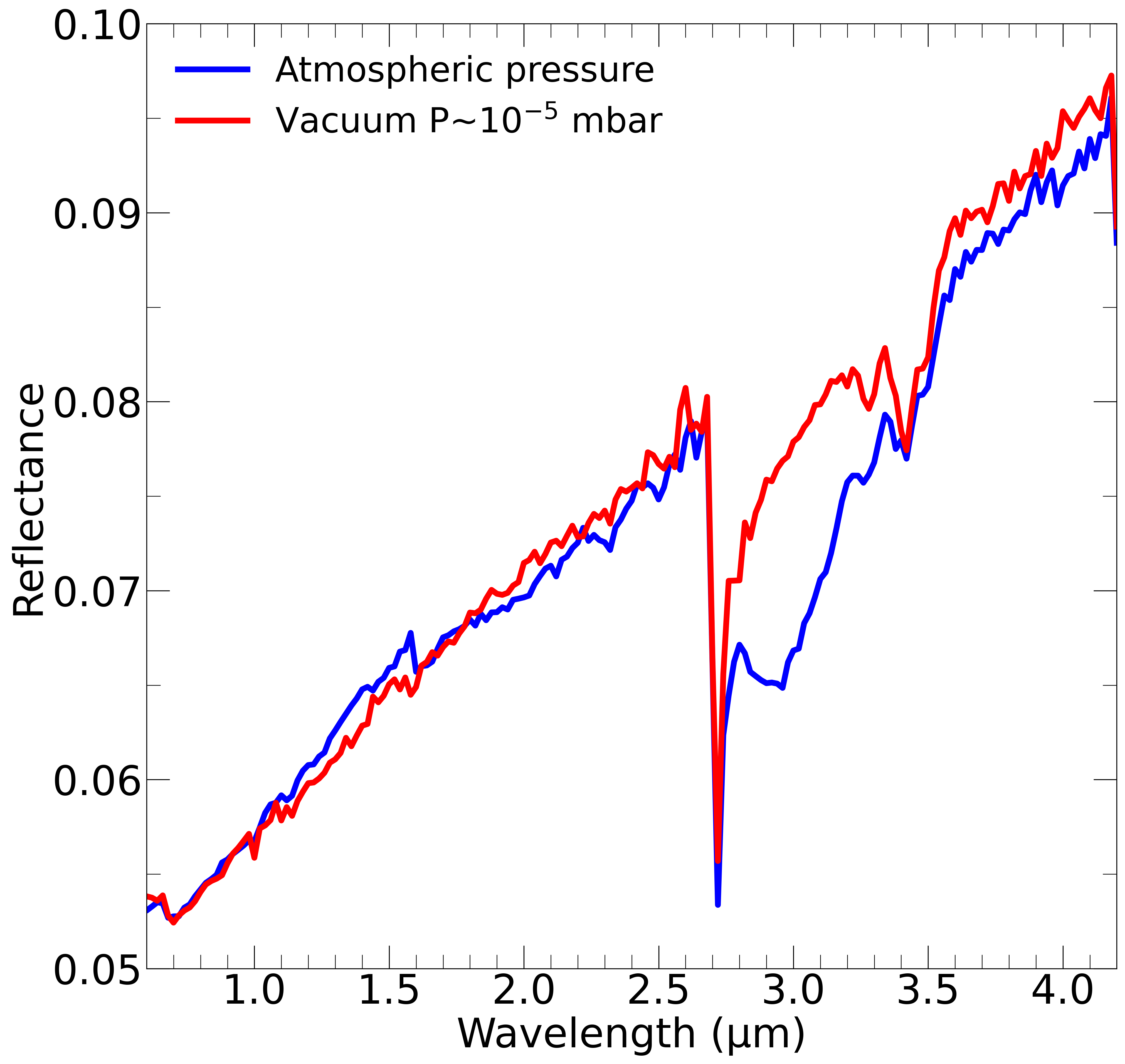}}
\caption{Comparison of two measurements made on the same sample (here, SIM-ATG-BIO-2) in vacuum (P$\sim$10$^{-5}$ mbar) and at atmospheric pressure. It shows that the 3 µm band for this sample is almost only linked to the adsorbed atmospheric water trap in the mixture powder.}
\label{fig:vacuum_effect}
\end{figure}

Vacuum measurements require the presence of a window at the top of the chamber. Measurements in different observation geometries cannot be performed in vacuum because the presence of a window could affect the results with refraction and multiple reflection. However, we evaluated the spectroscopic effect of vacuum by measuring some samples at i = 0$\degree$ and e = 30$\degree$ in the principal plane. Figure \ref{fig:vacuum_effect} shows a comparison between measurements taken at atmospheric pressure and at a pressure of 10$^{-5}$ mbar. We observed a complete removal of the 3 µm feature.

\printcredits

\bibliographystyle{cas-model2-names}

\bibliography{cas-refs}

\end{document}